\documentclass[traditabstract]{aa}  

\usepackage[varg]{txfonts}

\bibpunct{(}{)}{;}{a}{}{,} 

\DeclareUnicodeCharacter{2212}{-} 
\DeclareUnicodeCharacter{223C}{~} 

\begin{document}

\title{The high energy spectrum of Proxima Centauri simultaneously observed at X-ray and FUV wavelengths}

\author{B. Fuhrmeister\inst{\ref{inst1}}
\and A. Zisik\inst{\ref{inst1}}  
\and P. C. Schneider\inst{\ref{inst1}}
\and J. Robrade\inst{\ref{inst1}} 
\and J. H. H. M. Schmitt\inst{\ref{inst1}}
\and P. Predehl\inst{\ref{inst2}}
\and S. Czesla\inst{\ref{inst1}} 
\and K. France\inst{\ref{inst3}}
\and A. Garc\'ia Mu\~noz\inst{\ref{inst4}}}

\institute{Hamburger Sternwarte, Gojenbergsweg 112, 21029 Hamburg, Germany\\
  \email{bfuhrmeister@hs.uni-hamburg.de}\label{inst1}
  \and
   Max-Planck-Institut f\"ur extraterrestrische Physik, Gie{\ss}enbachstra{\ss}e 1, 85748 Garching, Germany\label{inst2}
  \and
   Laboratory for Atmospheric and Space Physics, University of Colorado Boulder, Boulder, CO 80303, USA\label{inst3}
  \and
   AIM, CEA, CNRS, Universit\'e Paris-Saclay, Universit\'e de Paris, 91191, Gif-sur-Yvette, France\label{inst4}
}

\date{Received <date> /accepted <date>}

\abstract{The M dwarf Proxima~Centauri, the Sun's
  closest stellar neighbour, is known to be magnetically active and it hosts a likely Earth-like planet
  in its habitable zone. High-energy radiation from the host star can significantly alter
  planetary atmospheres in close orbits. Frequent flaring may drive radiation-induced effects such as rapid
  atmospheric escape and photochemical changes. Therefore, understanding the characteristics of stellar radiation by
  understanding
  the properties of the emitting plasma is of paramount importance for a proper assessment of the conditions
  on Proxima~Centauri~b and exoplanets around M dwarfs in general.
  This work determines the temperature structure of the coronal and transition region plasma of
  Proxima Centauri from simultaneous X-ray and far-ultraviolet (FUV) observations. The differential emission
  measure distribution (DEM) was constructed for flaring and quiescent periods by analysing
  optically thin X-ray and FUV emission lines. Four X-ray observations of Proxima Centauri
  were conducted by the LETGS instrument on board of the \emph{Chandra} X-ray Observatory and four
  FUV observations were carried out using the STIS spectrograph on board
  the \emph{Hubble} Space Telescope. From the X-ray light curves, we determined
  a variation of the quiescent count rate by a factor of two within 20\% of the
  stellar rotation period.
  To obtain the DEM, 18 optically thin emission lines were analysed (12 X-ray and six FUV). The flare fluxes differ from the quiescence fluxes by factors of 4-20 (FUV) and 1-30 (X-ray).
  The temperature structure of the stellar corona and transition region was determined for
  both the quiescence and flaring state by fitting the DEM(T)  with Chebyshev polynomials
  for a temperature range $\log$T = 4.25 - 8. Compared to quiescence, the
  emission measure increases during flares for temperatures below 0.3\,MK (FUV dominated
  region) and beyond 3.6\,MK (X-ray dominated region).
  The reconstructed DEM
  shape provides acceptable line flux predictions compared to the measured values.
  Using the DEM we provide
      synthetic spectra  at 1-1700 \AA, which may be considered as representative for the
      high-energy irradiation  of Proxima~Cen~b during quiescent and flare periods. In future work these
      values can be used for planet atmosphere calculations which will ultimately provide 
      information about how habitable Proxima~Centauri~b is.}

\keywords{stars: individual: Proxima Centauri - stars: low-mass - stars: flare - stars: coronae - Ultraviolet: stars - X-rays: stars}

\titlerunning{The high energy spectrum of Proxima Centauri}

\maketitle

\section{Introduction}
At a distance of 1.3\,pc \citep{Gaia2018}, the red dwarf Proxima~Centauri, as the third member
of the $\alpha$ Centauri system, is the Sun's closest stellar neighbour. It has a spectral
type of dM5.5 \citep{Bessell1991}, corresponding to an effective temperature of 3040\,K 
\citep{Segransan2003}. The mass of Proxima~Centauri is  0.12\,$M_\odot$, its radius is 
0.15\,$R_\odot$ \citep{Segransan2003}, and its luminosity is 0.0015\,$L_\odot$ \citep{Doyle1990}. 
It is known to be magnetically active, as was
revealed for example in the X-ray regime by the \emph{Einstein} satellite \citep{Haisch1980}.
Despite its $\approx 50$ times smaller surface area compared to the Sun, Proxima~Centauri's quiescent
X-ray luminosity $L_X \approx (0.4 - 1.6)\cdot10^{27}$\,erg\,s$^{-1}$  is similar to the
Sun's \citep{Haisch1990}. During flares, bolometric energies of up to $\approx 10^{33.5}$\,erg
are released by the star \citep{Howard2018}, covering wavelengths from the X-ray to millimetre range
\citep{MacGregor2021}.
Nevertheless, Proxima~Centauri exhibits only a moderate
magnetic field
of B$f\approx\,600$\,G \citep{Reiners2008, Klein2021}, which is in line with its long rotation period
of about 83 days \citep{Benedikt1998}.

Proxima Centauri garnered increased interest, when
\citet{Anglada-Escude2016} announced the existence of a small planet (Proxima Centauri b)
with a minimum mass of 1.3\,$M_\oplus$, that is orbiting in the star's habitable zone at a
distance of 0.05\,AU (P$=11.2$ days). Questions began to rise as to whether the intrinsic
planetary properties as well as the stellar environment of Proxima Centauri allow for a habitable
planetary climate. 
In particular, the incident stellar radiation at short wavelengths -- which is X-ray, <100\,\AA\, and extreme ultraviolet (EUV), 100-912\,\AA\, -- can have adverse effects on the atmosphere of the planet at such close orbital
distances, especially during flares, when it is usually drastically increased. 
For example, high-energy radiation  can lead to the heating and ionisation of
atmospheric gas and, in
consequence, to thermal atmospheric escape \citep{Garcia2007, Murray-Clay2009, Koskinen2010, Salz2016}. At
longer wavelengths -- far ultraviolet (FUV), 912-1700\,$\AA$\, -- stellar radiation potentially induces
non-thermal chemistry (i.e. photo-chemistry), which can cause the loss of oceans and the
build-up of abiotic-produced O$_2$ and O$_3$ \citep{Luger2015, Loyd2018}. In contrast,
ozone layers
may be totally eroded by proton events \citep{Tilley2019}.

Stellar X-rays and ultraviolet (UV) photons are produced by the
hot plasma in stellar coronae and transition regions with a strong correlation of
X-ray luminosity to the magnetic field \citep{Reiners2012} and to the coronal temperature
\citep{Gudel2004a}. X-ray and UV radiation peaks
during flares, which are the result of magnetic re-connection at large atmospheric
heights \citep{Benz2017}. During such events, coronal temperatures reach values of up to 10 MK and more. Since
flare events are unpredictable and  cover only a fraction of the star's surface, the
plasma associated with flares spans a large range in temperature, at least from $10^4$
to $10^7$\,K \citep{Benz2017, Howard2020}.
However, X-ray and FUV telescopes cannot spatially resolve the surface of a
star other than the Sun, and therefore spectroscopic measurements
average information from multiple different plasma features. By determining the
differential emission measure (DEM) through spectroscopic analysis of X-ray and FUV
emission lines, the temperature structure of the emitting object can be reconstructed,
which we performed here for partially simultaneous X-ray data taken by \emph{Chandra}/LETGS
and FUV data taken by \emph{Hubble/STIS}. Since several flares occurred during the observations,
we took the opportunity to construct separate DEMs for flaring and the quasi-quiescent state of Proxima~Centauri.

Our paper is structured as follows. In Sect. \ref{sec:dem}, we give a short theoretical
introduction to the concept of the differential emission measure. In Sect.
\ref{sec:observations}, the observation information is listed. Sect. \ref{sec:data}
covers the timing as well as the spectral analysis of the X-ray and FUV data,
while the DEM fitting is described in
Sect. \ref{sec:construction}. Section \ref{sec:results}  contains our results and a
discussion of them, while Sect. \ref{sec:conclusion} presents a summary and conclusion.

\section{Differential emission measure}
\label{sec:dem}
Due to the mostly inhomogeneous nature of the stellar corona and transition region, measured
fluxes represent -- in reality -- the integrated emission of a large number of individual features
of the spatially unresolved, outer stellar atmosphere. The combined effect of  these different features, which are unknown to the
observer, can be parameterised by the DEM distribution
function $\xi(T)$.

\subsection{Computation of the DEM}
The DEM is a physical quantity that is related to the product of 
electron density $n_e$ and ion density $n_{ion}$,  as well as
the infinitesimal volume of plasma $dV$ associated with an infinitesimal range of 
temperature $1/dT$ \citep{Landi2008}. We further assume $n_{ion} \approx n_H \approx n_e$, with $n_H$
being the density of ionised hydrogen, and used the following for the DEM:
\begin{equation}
    \xi (T) = n_e^{2} \frac{dV}{dT}\,
,\end{equation}
with units in $\textrm{cm}^{-3} \textrm{K}^{-1}$. 
Another quantity often used in the literature is the volume emission measure ($EM$) defined as
\begin{equation}
EM = \int n_e n_H dV 
\end{equation}
with units in $\textrm{cm}^{-3}$. The DEM and the EM are related through 
\begin{equation}
EM_{\Delta T}(T) = \int_{T-\frac{\Delta T}{2}}^{T+\frac{\Delta T}{2}} \xi(T) dT\,.
\end{equation}

The DEM is an established description for
modelling plasma structures \citep[see, e.g.][]{Schmitt2004, Liefke2008, Duvvuri2021}.
It can be determined in many ways, almost all of which make use of
emission lines in the FUV  and X-ray ranges \citep{Landi2008}. When assuming the
plasma to be optically thin and in collisional ionisation equilibrium,
the observed emission line fluxes can be expressed as
\begin{equation}
    F(\lambda) = \frac{1}{4\pi d^2} \int_V G(T, \lambda, N_e)\; \xi(T)\; dT,
    \label{eq:flux_dem}
\end{equation}
where $d$ is the distance between the observer and the plasma, and $G(T,\lambda, N_e)$ is
the contribution function at wavelength $\lambda$ of all spectral lines. The
contribution function, which is strongly dependent on the plasma temperature $T$, is determined by
the atomic parameters of the spectral line and can be calculated using atomic databases.

Equation \ref{eq:flux_dem} constitutes a Fredholm equation of the first kind, and the inversion
of which
yields the DEM function $\xi (T)$. This type of integral inversion is ill-conditioned with no
unique solution for $\xi (T)$ unless one imposes additional constraints
\citep{Gudel2004a}. A formal discussion on this issue can be found in \citet{Craig1976}. For the
constraints applied here, readers can refer to Sect. \ref{sec:const}.

The basic integral equation can be solved numerically by replacing the integral operator
$G(T,\lambda, N_e)$ with a matrix operator. This approach is useful since the flux
$F(\lambda)$ is only measured at a finite number of points $\lambda_m$, and it is
therefore only possible to solve Eq. \ref{eq:flux_dem} for a discrete approximation to
$\xi(T)$. The matrix approximation may be obtained by discretising the temperature into
$n$ bins, so that
\begin{equation}
    F(\lambda_m) =  \frac{1}{4\pi d^2} \sum_{k=1}^n G(T_k, \lambda_m)\xi(T_k).
        \label{eq:DEM_matrix}
\end{equation}
Written in matrix form, the solution of
$\vec{\xi}$ is now obtained by the inversion of $\tens{G}$. This may be done in a variety of
ways, one of which  being the expansion of $\xi(T)$ 
in a series of orthogonal polynomials. Here we opted for
Chebyshev polynomials due to their favourable boundary conditions.

\subsection{Applied constraints to the DEM}\label{sec:const}
For the determination of the DEM, some constraints need to be imposed \citep{Schmitt2004},
with the most trivial one being
\begin{equation}
    \xi(T) \geq 0.
\end{equation}
The second constraint is 
\begin{equation}
    \xi(T_{max}) = 0
    \label{eq:tmax}
,\end{equation}
assuming a maximum temperature $T_{max}$ above which no emission measure is
present.
Finally, in the absence of any plausible physical model, the
distribution function $\xi(T)$ can be approximated by the sum of $N$ orthogonal polynomials
\citep{Schmitt2004}:
\begin{equation}
    \xi(T) = \sum_{i=0}^N a_i Ch_i(x),
\end{equation}
with $Ch_{i}(x)$ being Chebyshev polynomials of order $i$ and the dimensionless temperature
variable
$x$ defined in the closed interval [0,1]:
\begin{equation}
    x(T) = \frac{\text{log}(T) - \text{log}(T_{min})}{\text{log}(T_{max}) - \text{log}(T_{min})}.
\end{equation}
We determined the coefficients $a_i$  by a fit to the data. The used
boundary conditions are
\begin{equation}
    Ch_{2i}(0) = (-1)^{i} \qquad Ch_{2i+1}(0) = 0 \qquad Ch_{i}(1) = 1.
\end{equation}
Combining the third boundary condition with $\xi(1) = 0$ (equivalent to Eq. \ref{eq:tmax}),
the following relation was obtained for the Chebyshev coefficients $a_i$:
\begin{equation}
    \sum_{i=0}^{N} a_i = 0.
\end{equation}
Therefore, the number of independent coefficients $a_i$ is $N$ and the coefficient $a_0$ can
be
written as
\begin{equation}
    a_0 = -\sum_{i=1}^N a_i.
\end{equation}

\section{Simultaneous observations and data reduction}\label{sec:observations}

We obtained X-ray and FUV observations of Proxima~Centauri in 2017 with the \emph{Chandra}
and \emph{Hubble Space Telescope (HST)}, respectively, with the \emph{HST} data having been obtained
strictly simultaneously with the third out of the four X-ray observations.
This third observation is also
covered by Astrosat X-ray data and was analysed with a focus on the timing and multi-wavelength
behaviour by \citet{Lalitha2020}. Here the focus is on the construction of the DEM, and
the light curves were mainly used to identify flaring and quiescent episodes.

\subsection{FUV observations and data reduction}
\emph{HST} observed Proxima Centauri on 31 May 2017 during four consecutive orbits
(Prop. ID: 14860); for observation time details, readers can refer to Table~\ref{tab:obs}.
The combined exposure time of all four measurements was 12.28\,ks. All
observations were performed with the STIS/FUV MAMA detector \citep{Ward-Duong2021}. In combination with the echelle
grating E140M, the instrument provides spectra at a resolving power of $\sim 40\,000$ in the wavelength range 1140-1710\,\AA.
The aperture defining slits covered an area of 0.2$\times$0.2 arcsec on the sky as
recommended for point sources. All data were taken in \textit{TIME-TAG} mode, allowing us to
construct light curves in spectral windows which were defined a posteriori. Data were
downloaded from the \emph{Hubble Legacy Archive} and were reduced with \texttt{calstis}. 

\subsection{X-ray observations and data reduction}
The \emph{Chandra} X-ray observatory \citep{Weisskopf2002} observed Proxima Centauri between 15 May and 3 June 2017 four
times for a total of 175\,ks (PI: Predehl, Prop. ID: 18200754).
For details on the observation times, readers can refer to Table \ref{tab:obs}.
The combined exposure time of all four measurements was 165.93\,ks.
All observations were performed with the Low Energy Transmission Grating Spectrometer
(LETGS), which operates by using the Low Energy Transmission Grating (LETG) in conjunction
with the High Resolution Camera (HRC-S).  LETGS provides high-resolution spectroscopy
$(\lambda/\Delta\lambda > 1,000)$ between 50 and 160\,\AA\, and $\lambda/\Delta\lambda
\approx 20 \cdot \lambda$ at shorter wavelengths (3-50\,\AA). The full LETGS wavelength
range is 1.2 - 175\,\AA\, (0.07 - 10\,keV). The first order effective area varies between 1-25\,cm$^2$.
All X-ray data used in this study were downloaded from the \emph{Chandra} Data Archive and
reduced using CIAO 4.11. For standard processing and calibration, CALDB 4.8.3 was used.

\begin{table}[]
    \centering
    \caption{\emph{Chandra} LETGS and \emph{Hubble} STIS observation times}
    \footnotesize
    \begin{tabular}{lcc}
        \hline\hline
         Start date & Obs. ID & Dur. [ks]\\
         \hline
         \emph{Chandra} & & \\
         2017-05-15 23:27:56  & 20073 &  40\\
         2017-05-18 09:19:55  & 20080 &  52\\
         2017-05-31 16:24:33  & 19708 &  45\\
         2017-06-03 04:44:16  & 20084 &  30\\
         \hline
         \emph{Hubble} & & \\
         2017-05-31 21:19:36  & 201010&  2.5\\
         2017-05-31 22:32:21  & 201020&  3.2\\
         2017-06-01 00:07:21  & 201030&  3.2\\
         2017-06-01 01:43:12  & 201040&  3.2\\
         \hline
    \end{tabular}
    \label{tab:obs}
\end{table}


\section{Light curve analysis}\label{sec:data}
The following section provides an analysis of the X-ray and FUV light curves
obtained from the \emph{Chandra} LETGS and \emph{Hubble} STIS observations. The main
focus lies on the identification of flaring and
quiescent periods for the construction of a quiescent and a flaring DEM.

We classify Proxima Centauri's activity into  three different states, namely flaring, quiescent,
and intermediate state. Quiescent intervals are characterised by
low stochastic variability around a mean count rate. This quiescent count rate can change over time,
for example if the filling factor of hot material changes.
The flare identification process can be formalised, but ultimately remains
subjective, for example by setting threshold values. Given the low number of flares
in our case, the following criteria were applied manually:
The flaring periods were selected by identifying instantaneous (in seconds to
minutes) increases in the photon count rate by a minimum factor of three times the quiescent
rate.
The end of each flaring period was assigned when the count rate had decreased to approximately the
quiescent count rate. 
The remaining parts of the light curve -- which could not be assigned to quiescent or flaring state -- were classified as being in an intermediate state. For the FUV data, the two intermediate
time intervals seem to be located at the end of decay phases of flares, with one occurring before the observations started and the other being a secondary flare.
In the X-ray data, we identified a lot more
intermediate state time intervals, which are characterised by being above the
quiescent level, giving evidence of higher activity  without a
connection to clear flaring
signatures. To make sure that we are only dealing with the bona fide quiescent and
flaring state for the DEM construction, these intermediate intervals are excluded from the following analysis.
Flaring periods are marked in red in Figs.~\ref{fig:lc_h} and \ref{fig:lc_ch} in the \emph{Hubble} FUV
and \emph{Chandra} X-ray light curves, while phases of the quiescent state are marked in green.
Intermediate-type intervals are not marked. Flares are also numbered by roman numerals
for better reference.

\subsection{FUV light curves}\label{sec:lightcurve_fuv}
\begin{figure}
    \centering
    \includegraphics[width=1\columnwidth]{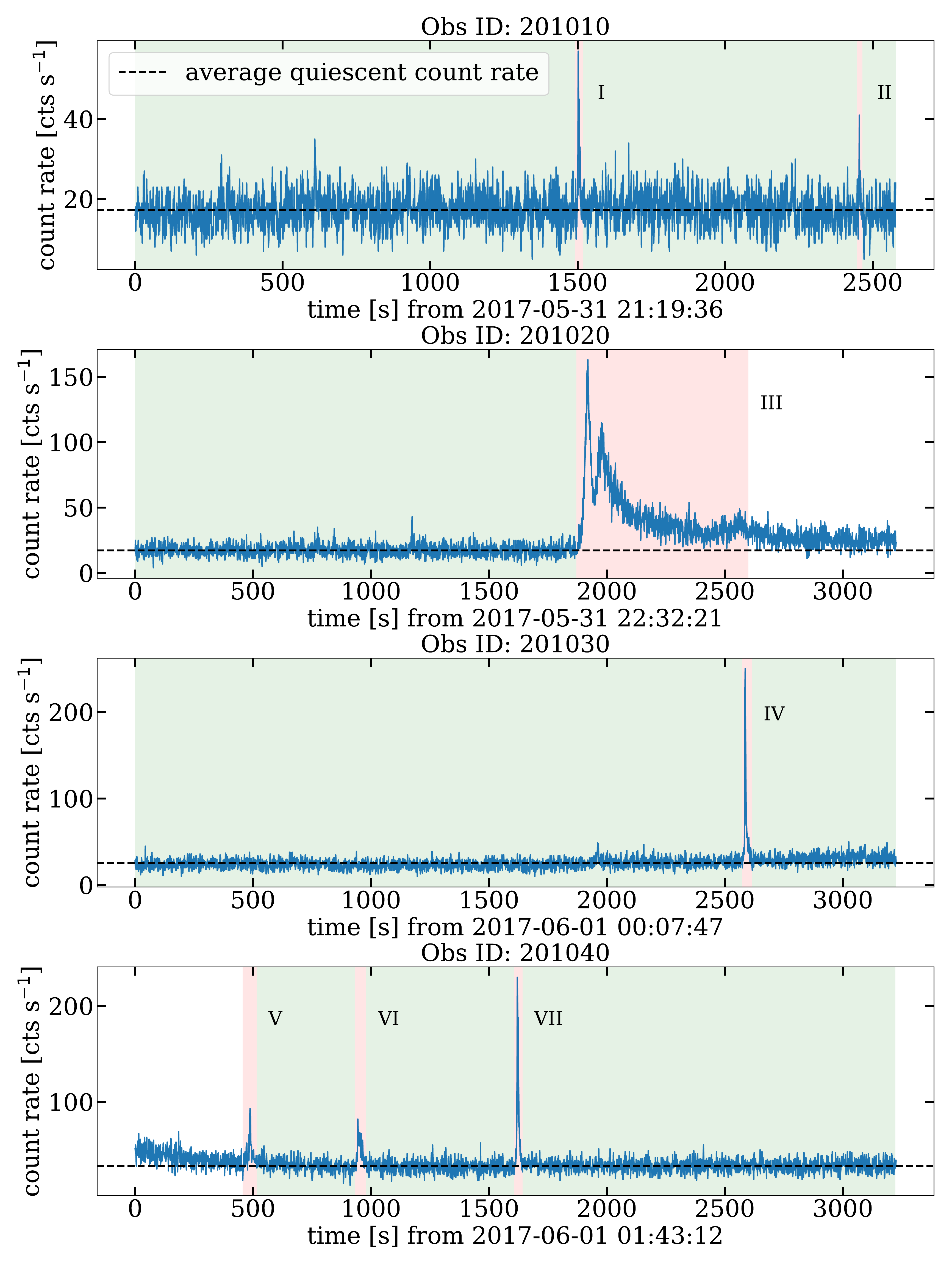}
    \caption{\label{fig:lc_h}Background-subtracted FUV light curves of \emph{Hubble}/STIS observations
      (blue line).
      The red shaded areas indicate flares, which are labelled with roman numerals.
      The green shaded areas show quiescent time
      intervals. The non-shaded intervals are classified as being in an 
	intermediate state.
      The dashed horizontal line indicates the average quiescent count rate in
      the individual observation runs.
      }
\end{figure}

The FUV light curves were extracted from the STIS/FUV-MAMA detector using the
\textit{python} module \textit{lightcurve}\footnote{\url{https://github.com/justincely/lightcurve/}}
and an integration range from 1100 to 1700 \AA. They are shown in
Fig.~\ref{fig:lc_h}. They display the background-subtracted photon count rates 
with a temporal binning of 1\,s. The FUV light curves show their typical
behaviour of a well-defined quiescent state, with the exception of pronounced flares.

The different observations for the first two runs (\texttt{201010} and \texttt{201020}) display
an average quiescent count rate of 17.32\,cts/s. In the third observation (\texttt{201030}),
we find a higher average quiescent count rate of 25.59\,cts/s, while the last observation (\texttt{201040})
has the highest quiescent count rate with 33.12\,cts/s. The higher average quiescent count
rates of the latter two observations can be explained by the higher overall activity state of
Proxima Centauri during that time, indicated by more frequent flaring and
by the continuous X-ray light curve displaying higher count rate levels, as well as more variability
during this time interval (see Fig. \ref{fig:lc_ch} for comparison). 
The quiescent time intervals of all four observations add up to 10.2\,ks with an average
count rate of 23.96\,cts/s, which corresponds to $L_{FUV,qu} = 1.4 \cdot 10^{26}$\,erg\,s$^{-1}$. 

\begin{table}[]
    \centering
    \caption{Flare properties of FUV and X-ray flares.}
    \footnotesize
    \begin{tabular}{lccc}
        \hline\hline
        flare & length & max.  & total  \\
              &        & count rate&   counts\\
               & [s]   & [cts/s]            &  [cts] \\
         \hline
         FUV {\sc I}  &  30 &  57 & 655 \\
         FUV {\sc II}  & 27 & 45  & 361 \\
         FUV {\sc III} &  729 & 163 & 33094 \\
         FUV {\sc IV}  & 41   & 251 & 2516\\
         FUV {\sc V}   & 47   & 97 &  2648\\
         FUV {\sc VI}  & 47  & 87  &  2277\\
         FUV {\sc VII} & 43  & 232 &  2574\\
         X-ray {\sc I} & 3000 & 0.79 &1508 \\
         X-ray {\sc II} & 880& 0.71 & 501\\
         X-ray {\sc III} & 850 & 1.04 & 438\\
         X-ray {\sc IV} & 1100 & 1.04 & 431\\
         X-ray {\sc V} & 1220 & 0.36 & 288\\
         \hline
    \end{tabular}
    \label{tab:flare}
\end{table}

The FUV light curve displays seven flare intervals as designated in Fig.~\ref{fig:lc_h}.
Their individual duration, maximum count rate, and total flare counts are listed
in Table~\ref{tab:flare}.
The total combined flaring time is 964\,s, with Flare {\sc III} contributing
$\sim$75 $\%$ of that time and also dominating the total flare counts. Moreover,
the morphology of Flare {\sc III} is different since it is the only flare with a notable duration and
also as it exhibits a secondary flare maximum.
This implies that the flare also dominates the DEM construction for the flaring time
interval. 
During flares, the photon count rate increases by factors of 2.4
to 10.5. The average count rate increase for the identified flares is a factor of 5.5 for the
flare maximum, while the average flare luminosity is $L_{UV,fl} = 7.7 \cdot 10^{26}$\,erg\,s$^{-1}$.

\subsection{X-ray light curves}\label{sec:lightcurve_xray}
\begin{figure}
    \centering
    \includegraphics[width=1\columnwidth]{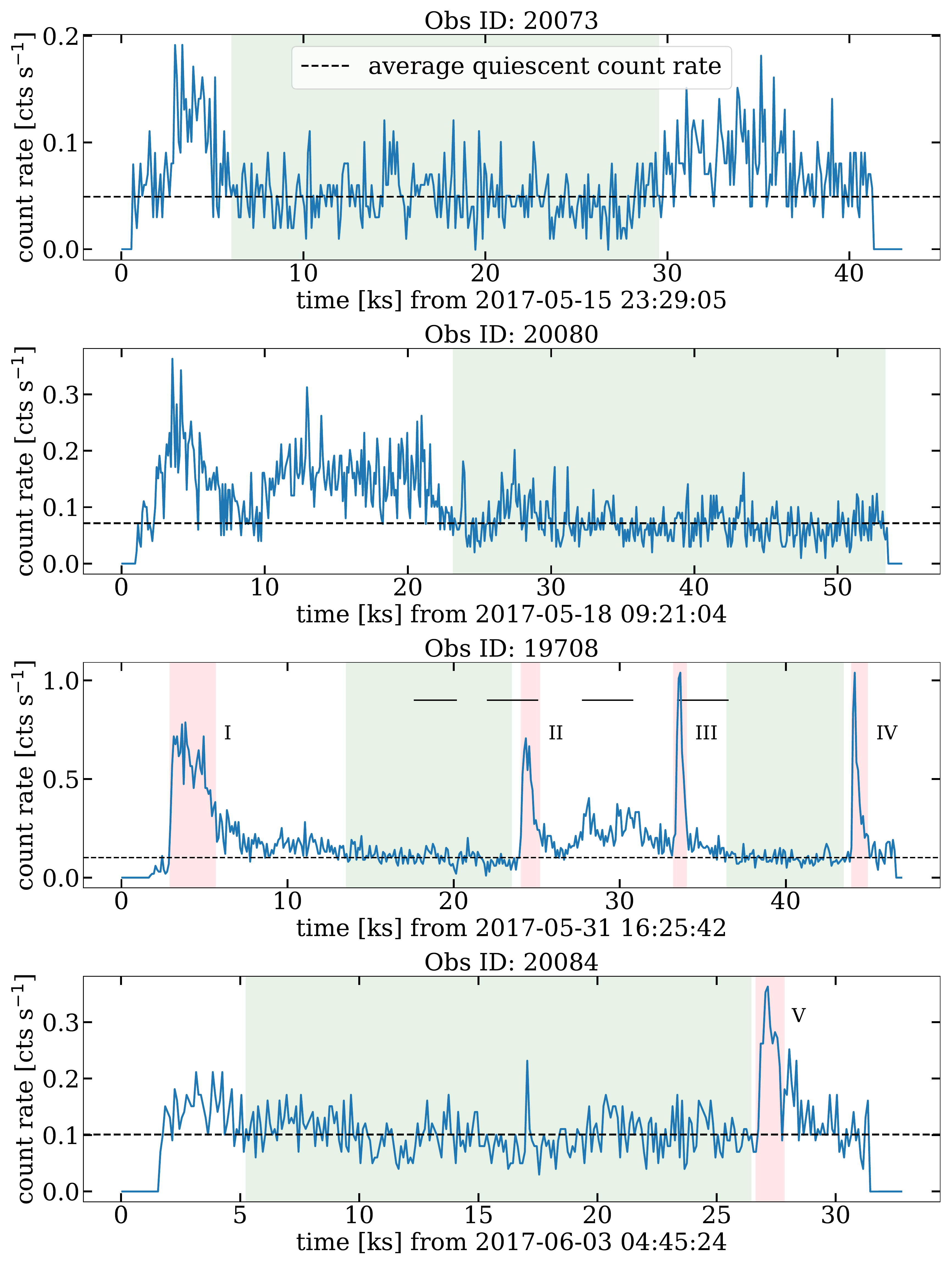}
        \caption{\label{fig:lc_ch}Background-subtracted X-ray light curves of \emph{Chandra}/LETGS 
        observations (blue line) constructed from the zeroth order.
      Flare and quiescent periods are shown in red and green shaded areas, respectively.
      Flares are labelled by roman numerals and the dashed horizontal lines indicate the
      quiescent average count rate for each observation. The black horizontal bars mark the
    time intervals of each of the four FUV observations.}
\end{figure}

The background-subtracted X-ray light curves extracted from the zeroth order of LETGS are shown in Fig. \ref{fig:lc_ch}.
The
temporal binning is 100\,s. During all four observations,
the X-ray activity level varies even on short time scales 
also exhibiting some sort of flickering outside the flares, which
is a rather typical behaviour of M dwarfs \citep{Robrade2005}. Examples of flickering can be
found in the time intervals classified as intermediate.

Observation \texttt{20073} displays mostly quiescent as well as slightly enhanced emission.
The quiescent period lasts for 23.5\,ks and has an average count rate of 0.049$\pm$0.002\,cts/s.
The light curve of observation \texttt{20080} shows a similar
picture: After a period of small variability with no clear
indications of a flare, a quiescent time interval follows. This lasts for 30.2\,ks with an average
0.072$\pm$0.002\,cts/s.
The light curve of observation \texttt{19708} displays two time intervals in which Proxima
Centauri is quiescent. Overall they last for 17\,ks with a combined average count rate of
0.102$\pm$0.003\,cts/s. Observation \texttt{20084} displays one long period of quiescence that
lasts for 21.2 ks with an average count rate of 0.101$\pm$0.002\,cts/s.
The quiescent time intervals of all four observations add up to 91.9\,ks with a total average
count rate of 0.077 cts/s, which equates to $L_{X,qu} = 6.6 \cdot 10^{26}$\,erg\,s$^{-1}$, assuming
the DEM peaks at $\log T = 6.6$ (the DEM is calculated in Sect. \ref{sec:results}, see Fig.~\ref{fig:DEM_qu}) and using an APEC model. This implies a rather low
activity state of Proxima Centauri during the observations as quiescent X-ray observations
range from 0.4 to 1.6 $\cdot 10^{27}$\,erg\,s$^{-1}$ \citep{Haisch1990}. 

Nevertheless, the quiescent count rates vary by a factor of 2 within a span of 17 days between
observations. Since this is only 20 $\%$ of the stellar rotational period, either a more active
region has rotated into view or the variation is due to the stochastic nature 
of stellar activity. 

The X-ray light curve displays five flare events as marked in Fig.~\ref{fig:lc_ch} with
their flaring properties given in Table~\ref{tab:flare}.
The total combined flaring time
is 7.03\,ks, with Flare {\sc I} dominating in time and total counts. This flare is not covered
by the FUV observations. 

During flares, the photon count rate increases by factors of 5 to 13.5. On average,
the count rate increases by a factor of 10 during flare maxima. The average flare luminosity
is $L_{X,fl} = 6.7 \cdot 10^{27}$\,erg\,s$^{-1}$. 

\subsection{Comparison of FUV and X-ray light curve}

\begin{figure}
    \centering
    \includegraphics[width=1\columnwidth]{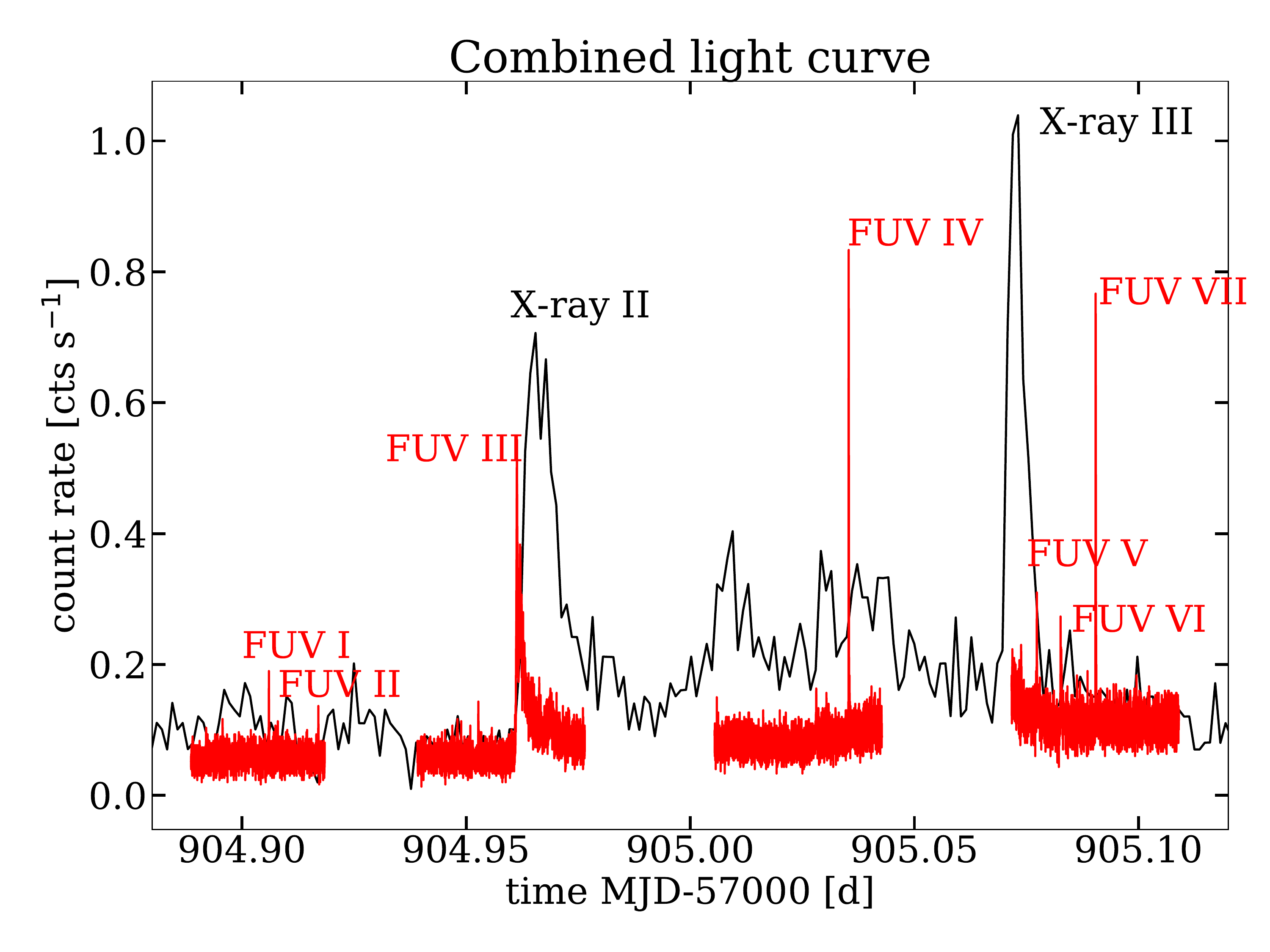}
    \caption{Combined light curve of \emph{Chandra} (black) and \emph{Hubble} (red)
      for the overlapping time period. Flare events are marked by their assigned roman
      numerals for X-ray flares in black and for FUV flares in red. The FUV light curve
    is scaled for convenience.}
    \label{fig:lc_comb}
\end{figure}

Figure~\ref{fig:lc_comb} shows both the \emph{Chandra} and \emph{Hubble} light curves in the
overlapping time interval. It can be noted that often distinct flares in the FUV wavelength
range do not need to have marked counterparts in the X-ray regime, as can be seen for FUV flare {\sc IV}
which occurs during X-ray flickering, without a larger flare associated with it or FUV flare
{\sc VII}, which is not even accompanied by X-ray flickering. This behaviour is possibly caused
by the better binning of the FUV data compared to the X-ray data and the shortness of
the FUV flares.\ Both of which lead to a veiling of the event in X-ray.

X-ray flare {\sc II} is
preceded by FUV flare {\sc III}, which is consistent with the Neupert effect because integration of
the FUV
light curve leads to a close resemblance of the X-ray flux increase at the start of the flare
until the X-ray peak is reached. X-ray flare {\sc III}
is associated with the onset of the fourth \emph{Hubble} observation, which shows a decrease
in count rate, also suggesting a preceding FUV flare event of an unknown
amplitude for X-ray flare {\sc III}.

\section{Construction of the DEM}\label{sec:construction}
\begin{figure}
    \centering
    \includegraphics[width=1\columnwidth]{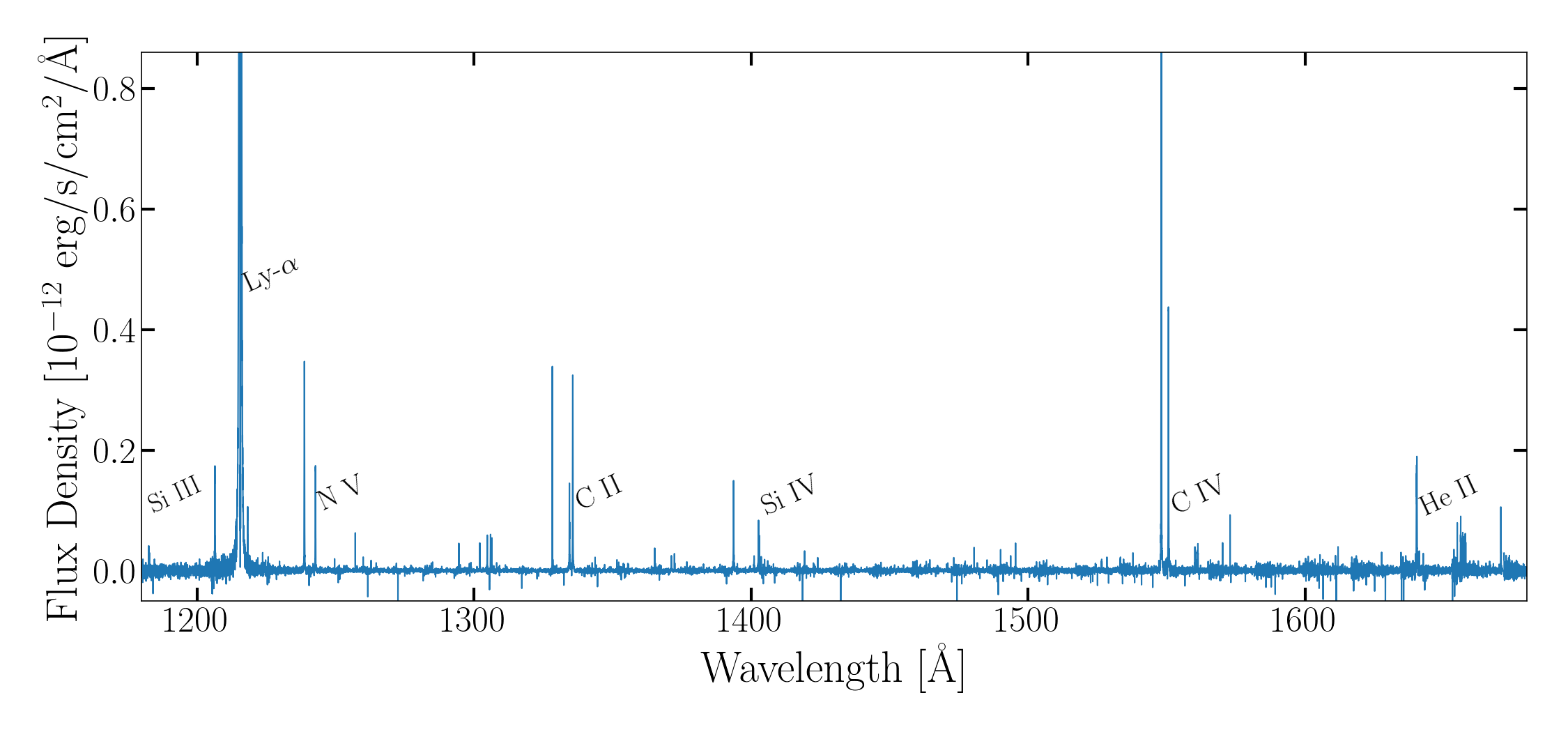}\\
        \includegraphics[width=1\columnwidth]{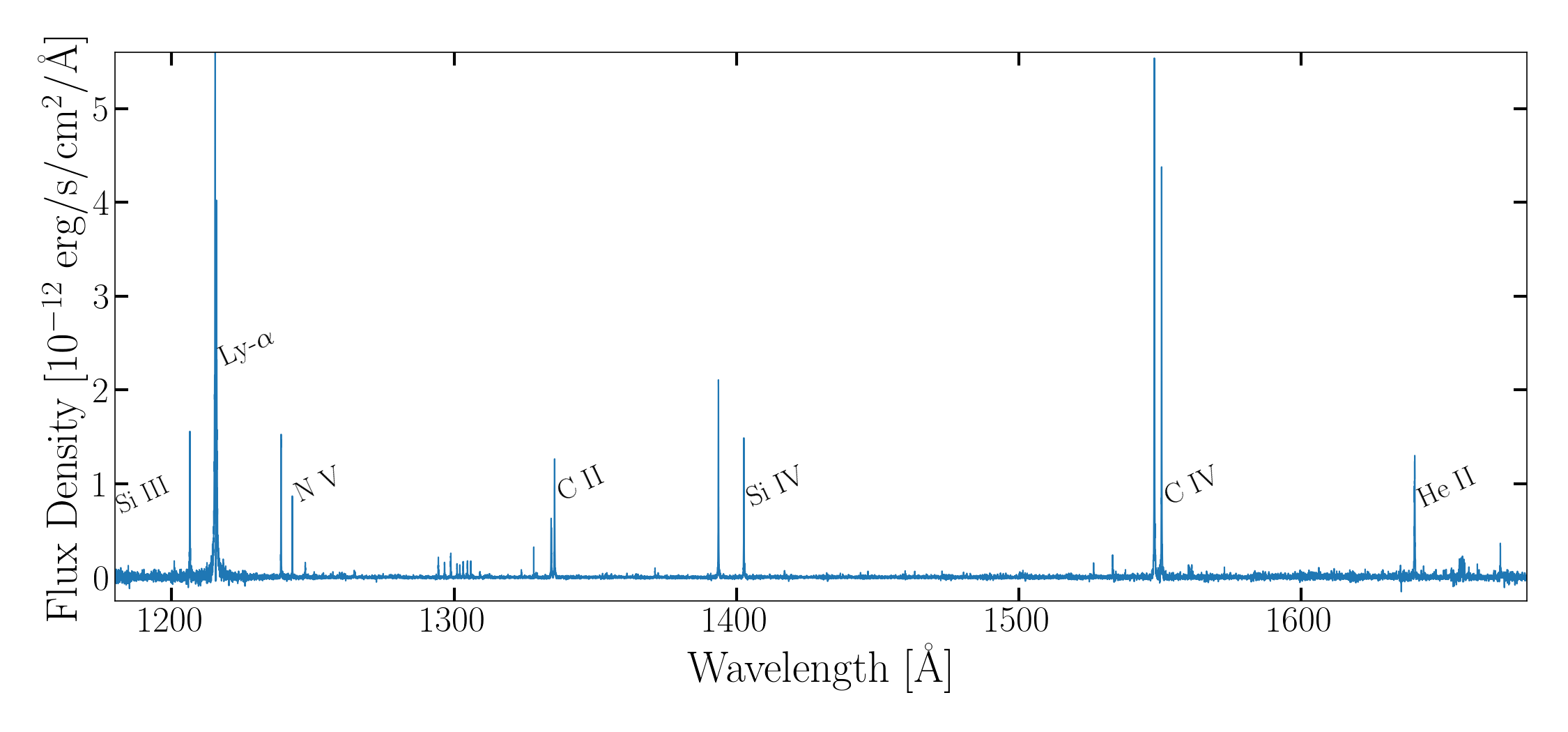}
    \caption{\emph{Hubble} FUV spectra extracted from the quiescent
        (top) and the flaring state (bottom). The most significant lines
        are labelled, even if they are not used for the further DEM analysis.}
    \label{fig:hubble_spec}
\end{figure}

With the quiescent and flaring periods identified in the previous step, spectra were  
generated by combining all flaring as well as quiescent periods for the FUV and X-ray
observations. 
In Fig.~\ref{fig:hubble_spec} we show the thusly constructed quiescent and flaring spectra of our FUV
data.
From these different spectra, line fluxes were obtained by fitting Voigt profiles to the 
emission line profiles in the FUV range and by integrating the spectrum for the
\emph{Chandra} X-ray lines because of their lower signal-to-noise ratio. 
Specifically, for the X-ray lines, the calculation was performed
by summing up observed photon counts within a
2$\sigma$ range, where $\sigma$ = 0.036 $\AA$ was estimated from a Gaussian fit to the 
strong \ion{O}{VIII} line at 18.97\,\AA. For each line, instrumental background and
continuum were also accounted for with the background counts
calculated in the wavelength intervals ranging from 3$\sigma$ to
9$\sigma$ of each line.

\begin{figure}
    \centering
    \includegraphics[width=1\columnwidth]{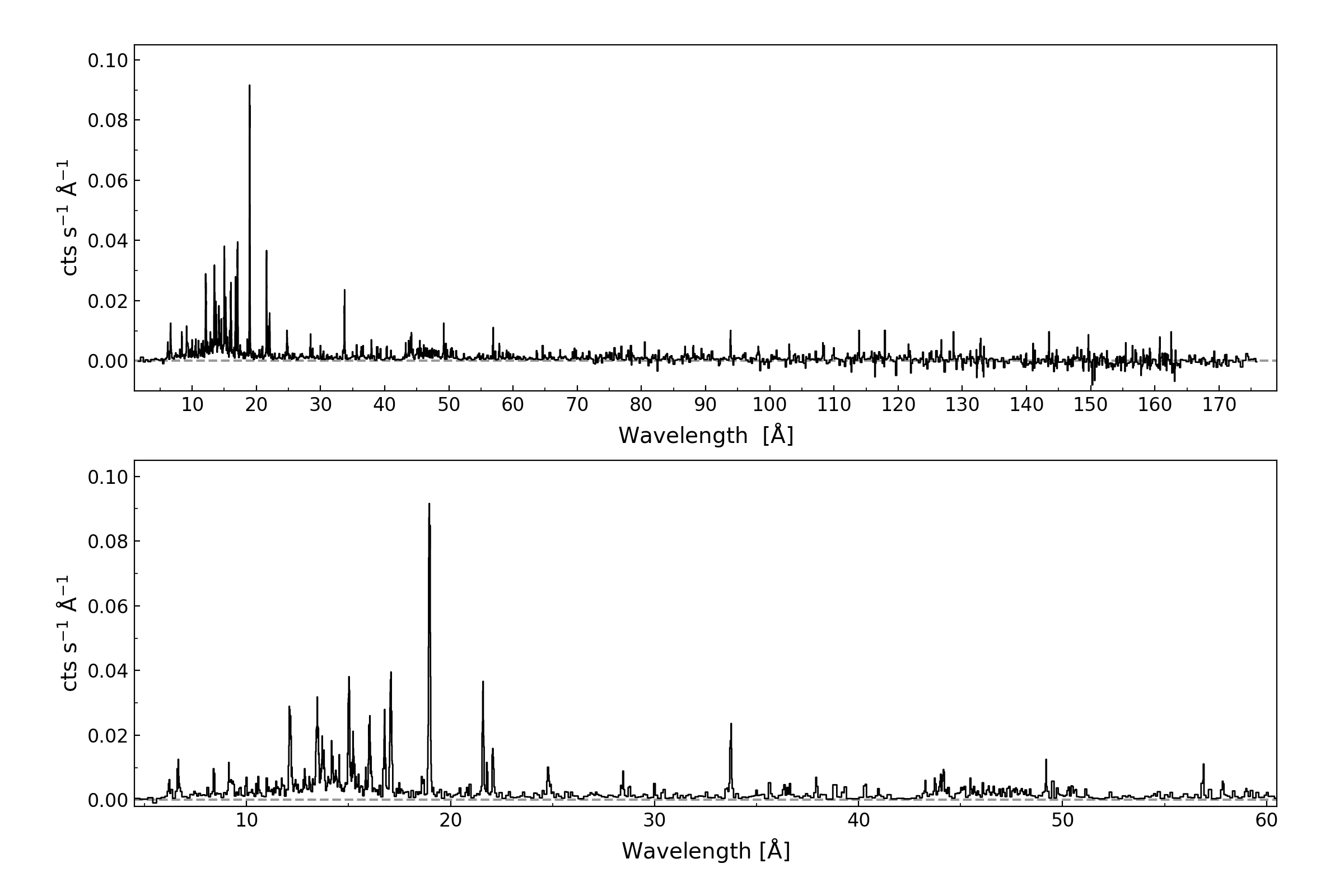}
    \caption{\emph{Chandra} X-ray spectrum for the whole observation (166\,ks) binned
        to minimum of abs(20) counts per bin (\emph{top}). The zoom (\emph{bottom})
        shows the  wavelength range used here.}
    \label{fig:xrayspec}
\end{figure}

To give an impression of the quality of the X-ray spectra, we show the X-ray spectrum
of the whole exposure (166\,ks) in Fig. \ref{fig:xrayspec}. The spectrum is in agreement
with a relative low emission measure at low temperatures generating X-ray
emission; this can be seen for example from the absent \ion{Fe}{ix} and \ion{Fe}{x}
lines at 171 and 174\,\AA, which have peak formation temperatures at $\log$T = 5.9 and 6.1,
respectively. We caution though that these two lines may also be undetectable due to
the low effective area of the instrument at these wavelengths ($\approx 1\,$cm$^2$, see
also Sect. \ref{sec:demres} for a more detailed discussion).

The  flux values for  25 emission lines
in the X-ray regime, which include ions of
Si, Mg, Ne, Fe, O, N, C, and Ni, were calculated for flare and quiescence periods. 
In the FUV, we found a total of 16 emission lines, which include ions of Si,
N, C, and He. Unfortunately the seven He lines are all blended so severely with wavelengths between
1640.33 and 1640.53\,\AA\  that we could not disentangle the lines. Moreover, the
line formation process for \ion{He}{ii} is not expected to be well-described by our assumption
of collisionally dominated plasma.
This led us to exclude these lines from any further analysis. 

Also, the quiescent FUV spectrum shows
some excess around the \ion{Fe}{XXI} line at 1354\,\AA. However, since this part of the spectrum 
is very noisy and the photon excess has a distinctly different shape than other lines (being much 
broader), we did not consider this \ion{Fe}{XXI} line in our analysis. We note, nevertheless,
that interpreting the excess as a genuine line flux would be roughly compatible with  other 
detected X-ray lines
having peak formation temperatures in the same temperature range.

\begin{figure}
    \centering
    \includegraphics[width=1\columnwidth]{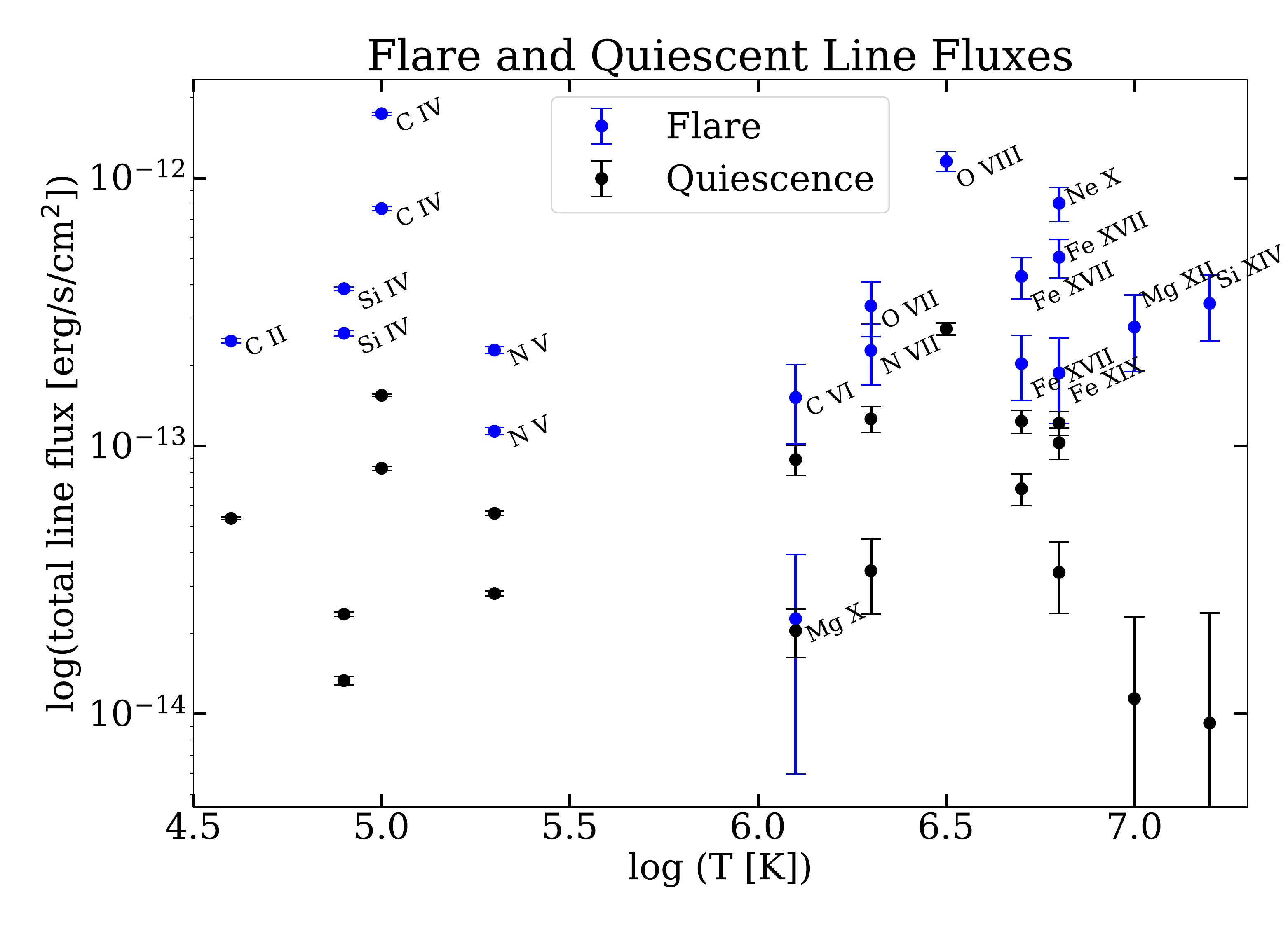}
    \caption{Total line fluxes of all X-ray and FUV emission lines plotted versus
      their peak formation temperatures during flares (blue) and quiescence (black).
      Both wavelength regimes are clearly distinct, with the FUV lines being
      below 10$^{5.5}$\,K and the X-ray lines being above 10$^{6.0}$\,K.}
    \label{fig:fluxes}
\end{figure}

From this preliminary set of emission lines, we identified suitable lines for the DEM fitting. 
For the FUV range, all line fluxes were measured with higher precision compared to the
X-ray line fluxes as can be seen in Fig.
\ref{fig:fluxes}. Unfortunately, two of the
lines show signs of self-absorption implying that these lines are not optically thin. We therefore decided to exclude
them from the further analysis.
The two lines are  a \ion{C}{II} line, whose line shape exhibits a distinct
double peak and a \ion{Si}{III} line, which
also shows signs of self-absorption during the flare.  
Moreover, we selected 15 X-ray emission lines with significant flux ($>3\sigma_{flux}),$
either in the quiescent or flaring spectra. From these, we excluded three
lines: one \ion{O}{VIII} line at 16.0 \AA,
which is a blend with another \ion{O}{VIII} line and a strong line of \ion{Fe}{XVIII}. Moreover, we
excluded one \ion{O}{VII} 
line at 22.1 \AA\ and one \ion{Ne}{IX} line at 13.7 \AA, which both are density sensitive, since they both are the forbidden line component 
of a He-like triplet \citep{Ness2002}. Thus we were left  with 12 X-ray lines for the
DEM fitting, using the same set of lines for a flaring and quiescent state, though some line fluxes are consistent with zero in one of the two states.
The measured properties of these lines can be found in the appendix
in Table~\ref{tab:xray_lines}  for the X-ray data and in Table~\ref{tab:fuv_lines} for the FUV lines.

The total line fluxes are shown in Fig.~\ref{fig:fluxes}, where they are plotted against their peak 
formation temperatures as listed in the APED database\footnote{\url{http://www.atomdb.org/}} \citep{APED}.
The peak formation temperature 
is the temperature at which the emissivity of an atomic line is maximal.

A clear distinction between FUV and X-ray lines can be seen in Fig. \ref{fig:fluxes}. 
The FUV lines are most
efficiently produced at temperatures between $\log$ T = 4.6 - 5.3 (40\,000 - 200\,000\,K), while the X-ray lines
are emitted in the temperature range $\log$ T = 6.1 - 7.2 (1.2 - 15.8 MK). In our data,  
no lines that formed in the temperature range $\log$ T = 5.4 - 6.0 (250\,000 - 1\,000\,000\,K) were measured. Such lines  
would be emitted mostly in the extreme ultraviolet (EUV, 100 - 1200\,\AA), of which only a
fraction is covered by \emph{Chandra}  and being hampered by 
low sensitivity in this wavelength region. 
We therefore chose the definition range for our DEM calculations to start from
$\log$(T$_{min}$) = 4.25 since our data are largely insensitive to cooler plasma. While this lower temperature end comes without 
any further constraints, we set
$\log$(T$_{max})=8.0$, with the constraint that the DEM(T$_{max}$)=0, as
detailed in Sect. \ref{sec:const}, since we expect no relevant emission above this
temperature.

The strongest emission line in the FUV in our data is \ion{C}{IV}. The strongest X-ray lines are
\ion{O}{VIII}, \ion{O}{VII}, and \ion{Fe}{XVII}. The differences in the errors of X-ray and FUV 
fluxes are caused by the lower sensitivity of \emph{Chandra} in comparison to 
\emph{Hubble}/STIS, which leads to comparably few counts in the X-ray range resulting in
larger statistical errors. 

For the construction of the DEM from this line set, in principle, the solution with the lowest order polynomial which still provides an acceptable
fit to the data should be chosen. In the case of this study, polynomials of order five are used for the quiescent data and those of order six for flaring data.
The fit of the quiescent DEM function was conducted while varying the
elemental abundances of the coronal plasma with the Fe abundance fixed to one.
For the flaring DEM fit, we first fixed the abundances to the ones obtained
for the quiescent DEM and fitted the Chebyshev polynomial coefficients, then in
a second step we fixed the coefficients and fitted the abundances.
Further we calculated errors for the DEM fit as follows: To calculate
the errors of the Chebyshev polynomial coefficients, we kept the abundances fixed and chose random flux values within the obtained errors and we re-calculated
the fit 100 times. To calculate the errors of the abundances, we  in
turn kept the coefficients of the Chebyshev polynomial fixed. Although we chose this
method for its computational robustness, we caution, nevertheless, 
that changes in the DEM or abundances
may compensate for each other and still yield similar fluxes. Therefore, this method may underestimate
the errors.



\section{Results}\label{sec:results}

\subsection{Elemental abundances}
The best-fit abundances obtained from the DEM fitting process described in Sect. \ref{sec:construction}
are listed in Tab. \ref{tab:abund}
with respect to solar abundance as given by \citet{AndersGrevesse1989}. 
Although we derived abundances that are in detail different from those by \citet{Gudel2004b} and \citet{Fuhrmeister2011}, our results are
overall similar. For an active star such as Proxima~Centauri, an inverse first ionisation potential (FIP)
effect-like abundance pattern is expected, that is elements of low FIP are depleted compared to elements
of high FIP \citep{Audard2003}. While \citet{Fuhrmeister2011} found a tentative inverse FIP pattern
during the quiescence state for Proxima~Centauri, \citet{Gudel2004b} found a rather flat abundance pattern
close to solar photospheric values. Our abundance pattern is inconclusive concerning a FIP or inverse FIP effect,
but it is also  relatively close to solar photospheric values. This may be explained by the rather inactive
state of Proxima~Centauri during our observations, as seen by the low X-ray and FUV fluxes (as discussed
in Sects. \ref{sec:lightcurve_fuv} and \ref{sec:lightcurve_xray}). Also, 
most of our flaring
abundances are quite similar to the quiescent values, which is as expected since the observed flaring activity does not
include large flares.

\citet{Gudel2004b} also used the solar abundances by \citet{AndersGrevesse1989}, except for
Fe, where they used \citet{GrevesseSauval1999}. Here, we
compare our abundance analysis to their results in more detail. 
We recalculated their abundancy ratios to the abundancy stated by \citet{AndersGrevesse1989}
and found no difference for the given accuracy, except for Fe/H=0.50 instead of their original value of 0.51. Therefore, we provide their abundance ratios in 
Table~\ref{tab:abund}. We opted for a normalisation with respect to Fe
since the largest number of used lines in the X-ray regime originate from Fe.
Using this normalisation to an Fe abundance of 1.0, all 
given values also represent abundance ratios relative to Fe.
Thus the abundance ratios are in agreement with those from \citet{Gudel2004b}, except for O/Fe, which is
higher in our data. 

\begin{figure}
    \centering
    \includegraphics[width=1\columnwidth]{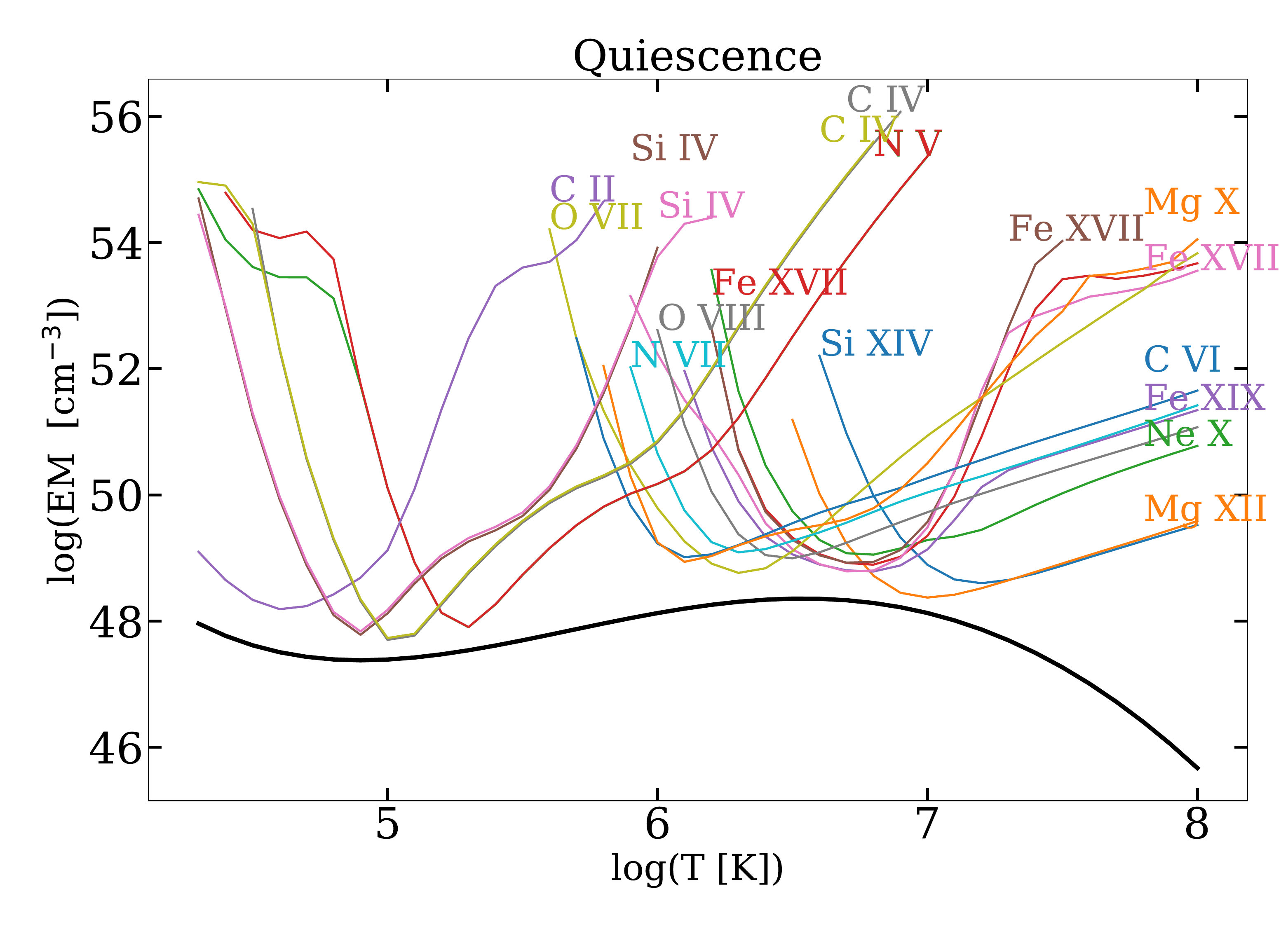}\\
    \includegraphics[width=1\columnwidth]{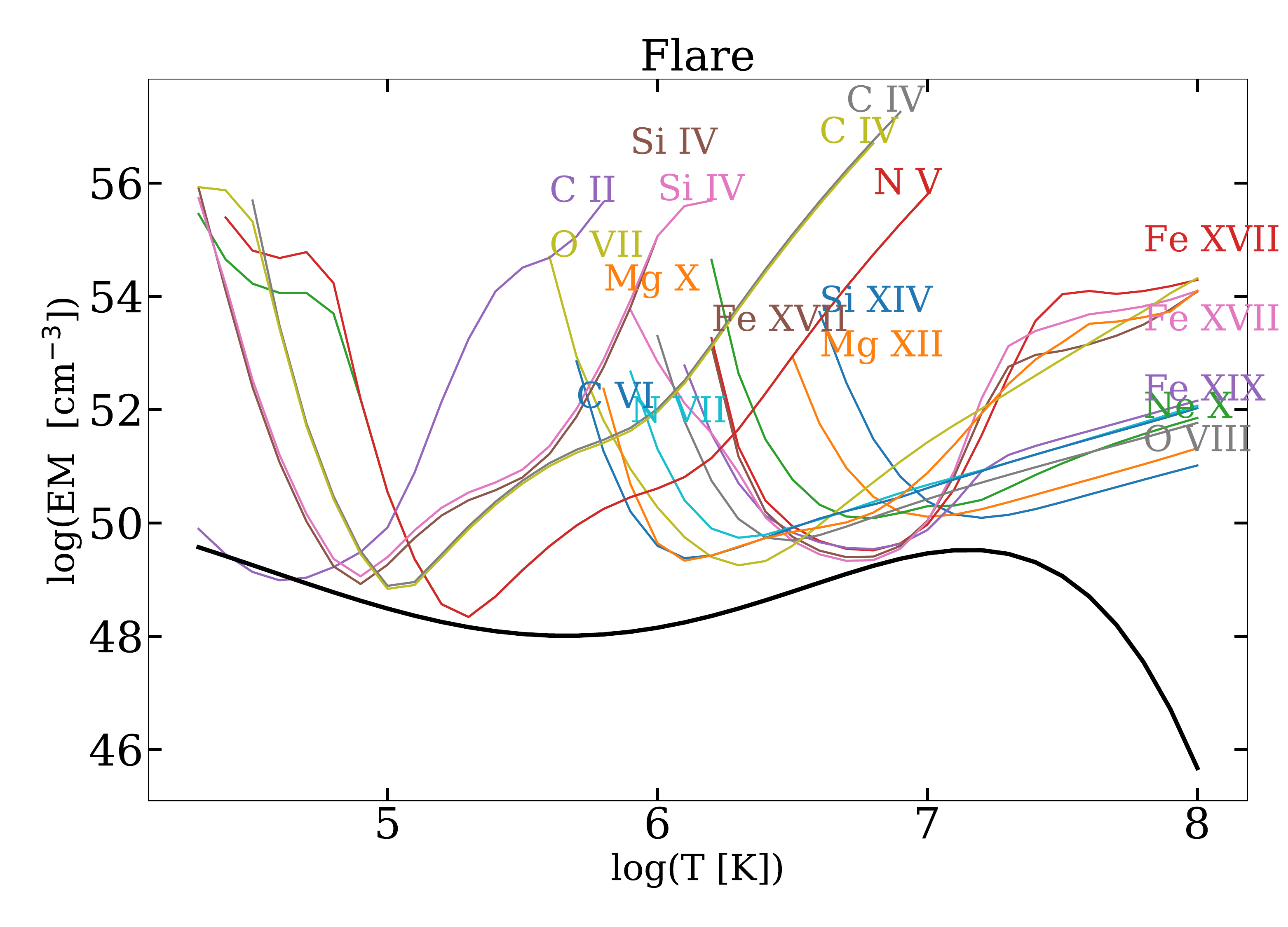}
    \caption{Discretised emission measure distribution of Proxima Centauri (black line)
      during quiescence
        (\emph{top}) and during combined flare periods (\emph{bottom}).
      As assumed by the fitting model, the emission measure vanishes at
      high temperatures. Coloured lines represent the observed line fluxes divided by their
      contribution function. Their minima are at the peak formation temperatures of the
      respective line.}
    \label{fig:DEM_qu}
\end{figure}

\begin{table}[]
    \centering
    \caption{Elemental abundances relative to solar photospheric values.}
    \begin{tabular}{cccc}
        \hline\hline
            Ion & \multicolumn{2}{c}{Relative Abundance} &\citet{Gudel2004b}  \\
         \hline
            & Quiescence & Flare & Quiescence\\
            C/Fe & 1.45 $\pm$ 0.31 & 1.07$\pm$ 0.18 & 1.9$\pm$0.7\\
            N/Fe & 1.19 $\pm$ 0.23 & 1.77$\pm$ 0.47 & 1.5$\pm$0.6\\
            O/Fe & 1.44 $\pm$ 0.33 & 1.23$\pm$ 0.30 & 0.6$\pm$0.2\\
            Ne/Fe& 2.06 $\pm$ 0.61 & 1.34$\pm$ 0.36 & 1.6$\pm$0.6\\
            Mg/Fe & 3.5 $\pm$ 0.46 & 1.54$\pm$ 0.91 & 2.1$\pm$0.9\\
            Si/Fe & 1.67 $\pm$0.75 & 1.99$\pm$ 0.21 & 2.1$\pm$0.8\\
            Fe/H & 1.0 & 1.0 &0.50$\pm$0.15\\
          \hline
    \end{tabular}
    \label{tab:abund}
\end{table}

\subsection{DEM}\label{sec:demres}
 
The results of the DEM fitting process are shown in Fig. \ref{fig:DEM_qu}
for quiescence (top panel) and the flaring (bottom panel) period, where the black line represents the
discretised emission measure  of Proxima Centauri with a logarithmic binning of
$\text{log}(T_{i+1}) - \text{log}(T_{i}) = \text{log}(T_{i+1}/T_i) = 0.1$. The coloured lines
represent the measured line fluxes of all emission lines that were selected for the DEM
fitting, divided by their contribution function. These curves are sharply peaked in temperature
with their minima being at the respective peak formation temperature of the emitting line. Only some
curves, especially of X-ray lines (e.g. \ion {O}{VIII}), are somewhat flat towards higher temperatures. These irregularities are caused by the calculation of the
contribution function with the help of atomic databases. If ions are listed in the database at similar
wavelengths to the observed emission line, they contribute
to the emission depending on their theoretical
relative flux, but often at different temperatures. Since their contribution is negligible in comparison
to the observed lines, this behaviour did not compromise the fitting of the DEM curves.

The binned flare EM peaks around $\log$ T = 7.2 (15.8 MK),
while a smaller secondary peak can be seen at
the low temperature end, which indicates enhanced emission of UV
photons during flares. The quiescence EM expectedly peaks at a lower plasma temperature,
$\log$ T = 6.3 (2.0 MK), which is still within the X-ray dominated
temperature range. Beyond a local broad minimum of the quiescence EM  
around $\log$ T = 4.8 (63\,000\,K),
the EM increases again to the low temperature end.
These results roughly match the
calculations of \citet{Gudel2004b}, who observed Proxima Centauri in X-rays using the X-ray satellite
\emph{XMM-Newton}
and found the EM distribution during low-level episodes to be dominated by
plasma at $\log$ T = 6.48 (3.0 MK). For an observed flare, they found the
EM to peak at $\log$ T = 7.18 - 7.30 (15.1 - 20 MK). Since they
lacked FUV data, their EM fitting starts at 1 MK.

\begin{figure}
    \centering
    \includegraphics[width=1\columnwidth]{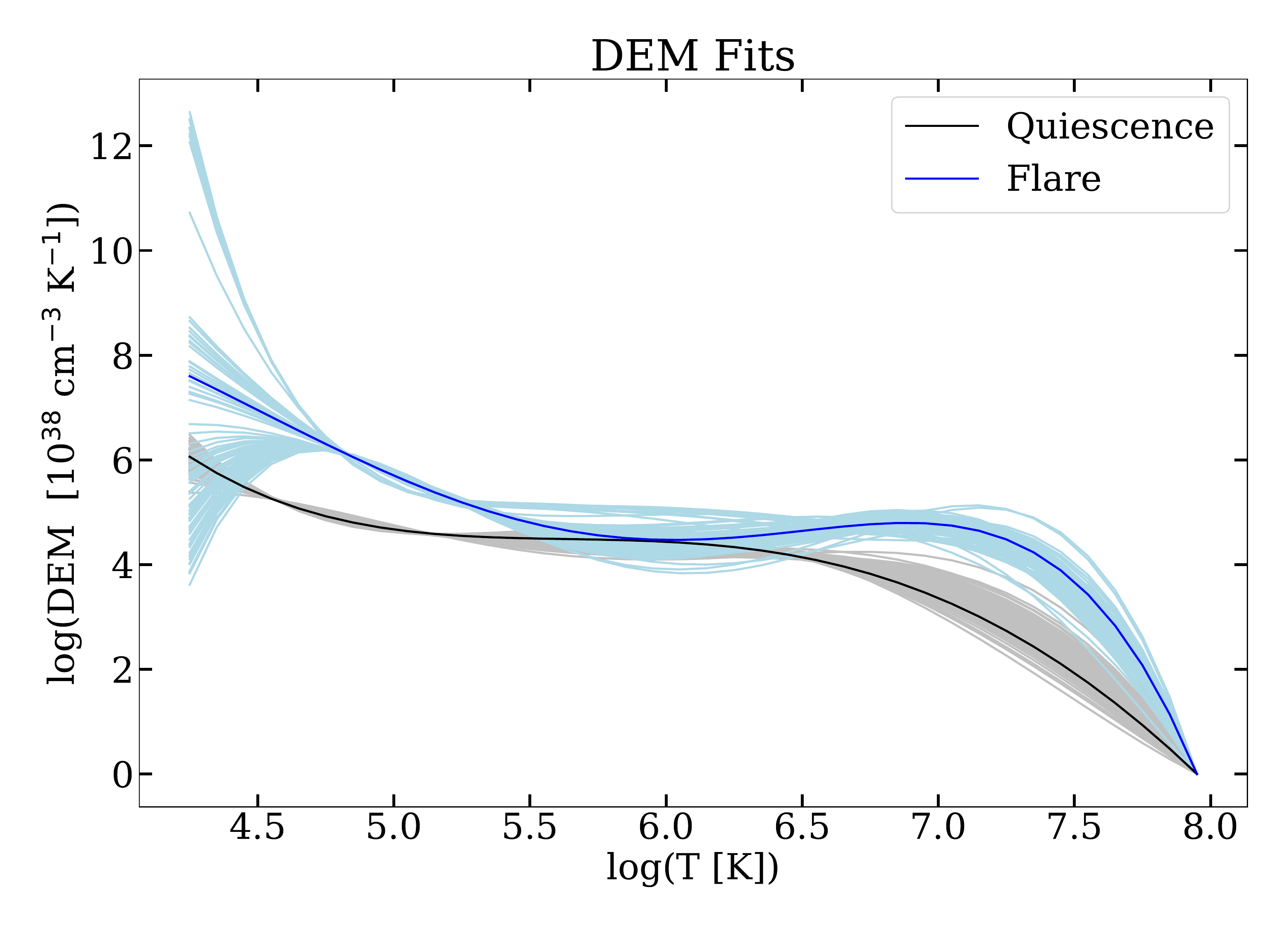}
    \caption{DEM distribution of Proxima Centauri for 
        quiescent (black) and flaring (blue) periods with computed errors.}
    \label{fig:DEMs}
\end{figure}

In Fig. \ref{fig:DEMs} the DEM distributions during 
flares and quiescence are shown together with the obtained errors.
The flare-DEM exhibits significantly higher values than the quiescence-DEM for nearly every
temperature besides the temperature region around $\log$ T=5.8. Those temperatures are
not well-constrained in our data 
because lines with formation temperatures in that range are neither 
significantly detected in the quiescent nor flare state. Although the formation temperature
of the \ion{Fe}{IX} line at 171\,\AA\, 
and the \ion{Fe}{X} line at 174\,\AA\, is in the considered regime, these lines fall 
in a detector region with very poor sensitivity (effective area of $\approx1\,$cm$^2$) and a high 
background, making meaningful conclusions challenging. In any case, their nominal 
upper limits are in contradiction with the predicted fluxes based on the 
reconstructed DEM and they would require somewhat lower (D)EM values around roughly $\log$ T=5.8. 
Abundance effects are likely insufficient to explain this discrepancy between the ``low temperature'' Fe lines and 
the DEM prediction because  Fe lines with a higher formation temperature 
are generally compatible with lines of other ions. 
Therefore, any DEM respecting the low temperature Fe-line upper limits
would greatly underestimate several other, clearly detected lines (chiefly \ion{C}{vi} and 
\ion{O}{vii}),
so that the inclusion of these low temperature Fe lines into the fit has only a marginal effect
on the resulting DEM.  
Therefore, we conclude that 
the DEM around $\log T\approx5.8$ may be slightly, but not greatly, lower than what was reconstructed. 

\begin{figure}
    \centering
    \includegraphics[width=1\columnwidth]{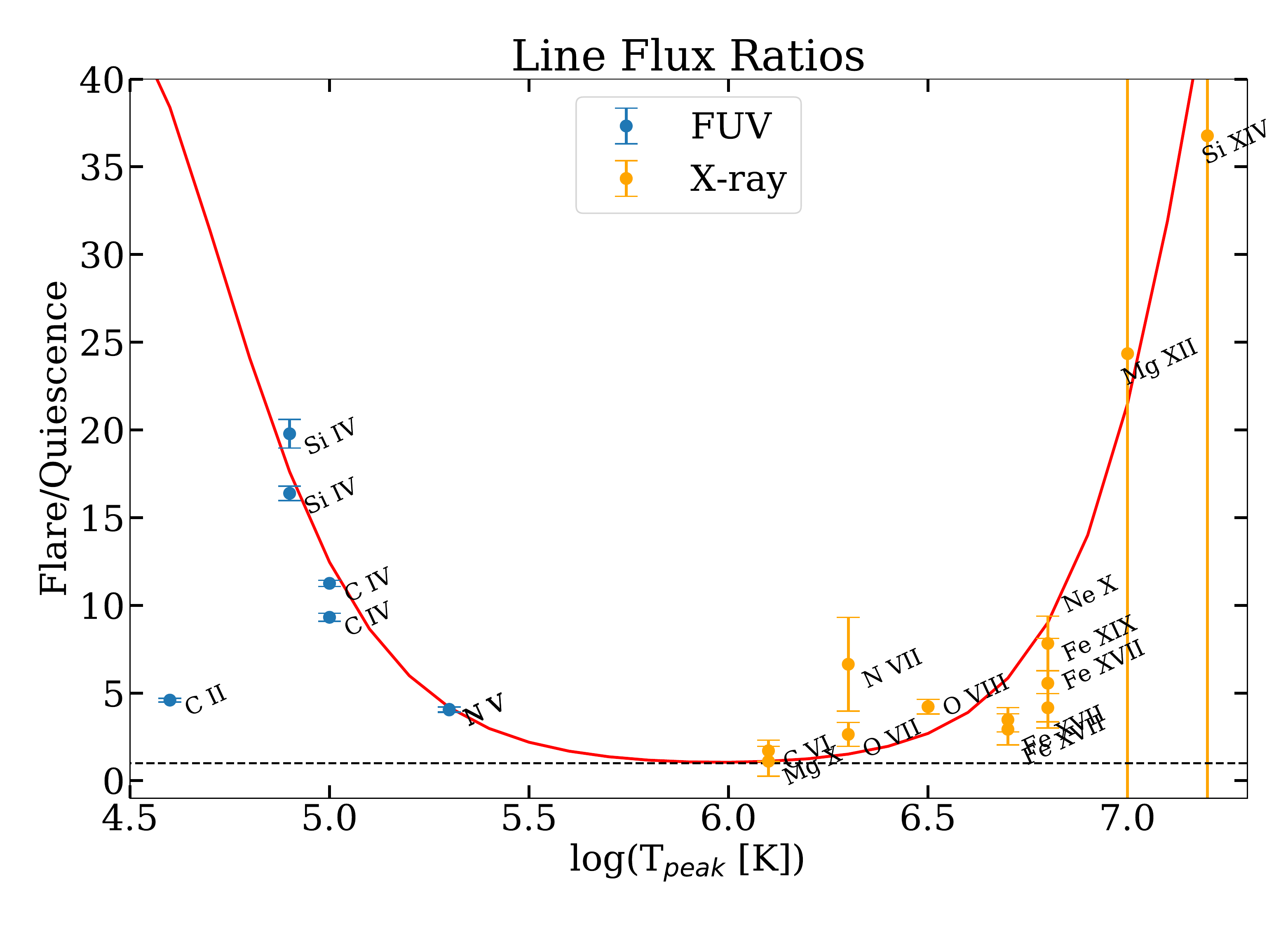}
    \caption{Flare to quiescence line flux ratios at their respective peak formation temperatures.
      The dashed black line indicates 1:1 ratios. An increase in the ratio
        is seen to lower
        and higher temperatures, respectively, with an exception for
        \ion{C}{II}.
        The ratio of the flare and quiescence DEM(T) functions (red line)
      reproduces this behaviour roughly.}
    \label{fig:ratios}
\end{figure}

The ratio between flare and quiescent fluxes (see Fig.~\ref{fig:ratios}) as a function of the peak 
formation temperature indicates that the low ($\log T\lesssim5$) and high ($\log T\gtrsim6.5$) 
temperature plasma components have the largest ratio between flare and quiescent fluxes, while 
lines that formed in an intermediate temperature range ($5.3<\log T<6.5$) exhibit lower ratios.
This behaviour is captured by the reconstructed DEMs; although, we admit that we do not have
significantly detected lines between $\log T=5.5$ and $\log T=6.0$, thus we cannot place 
firm conclusions on the evolution of this intermediate temperature plasma component.

While the quiescent DEM is rather well-defined, the flare-DEM exhibits some 
uncertainty at the low temperature end. In particular, we deem
DEMs that show a decrease towards low temperatures physically insensible, since it is 
known that the DEMs increase to lower (chromospheric and photospheric) values; such
shapes are likely only mathematically preferred solutions, caused by the assumption of (low order)
Chebyshev polynomials.

\begin{figure}
    \centering
    \includegraphics[width=1\columnwidth]{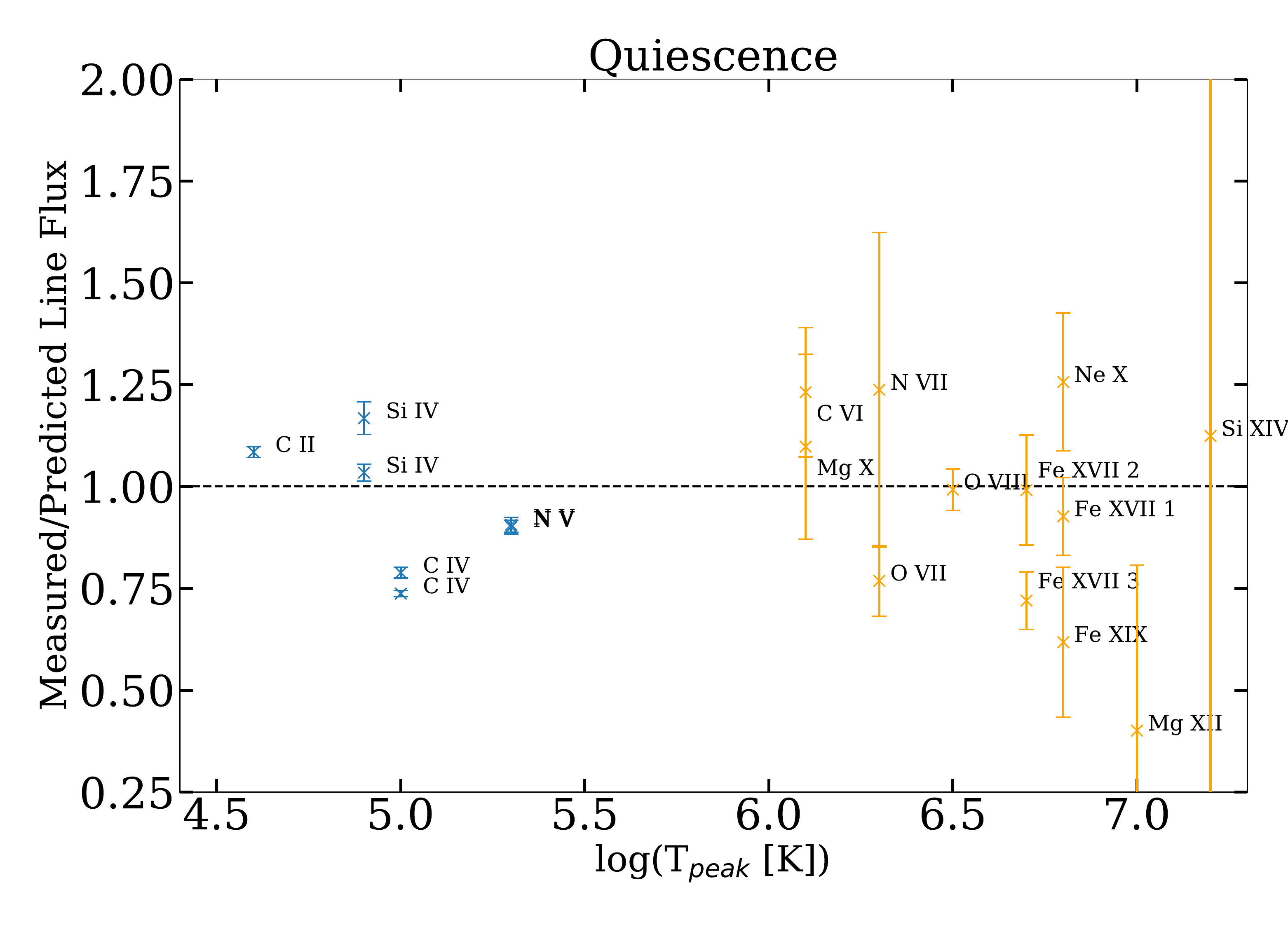}\\
    \includegraphics[width=1\columnwidth]{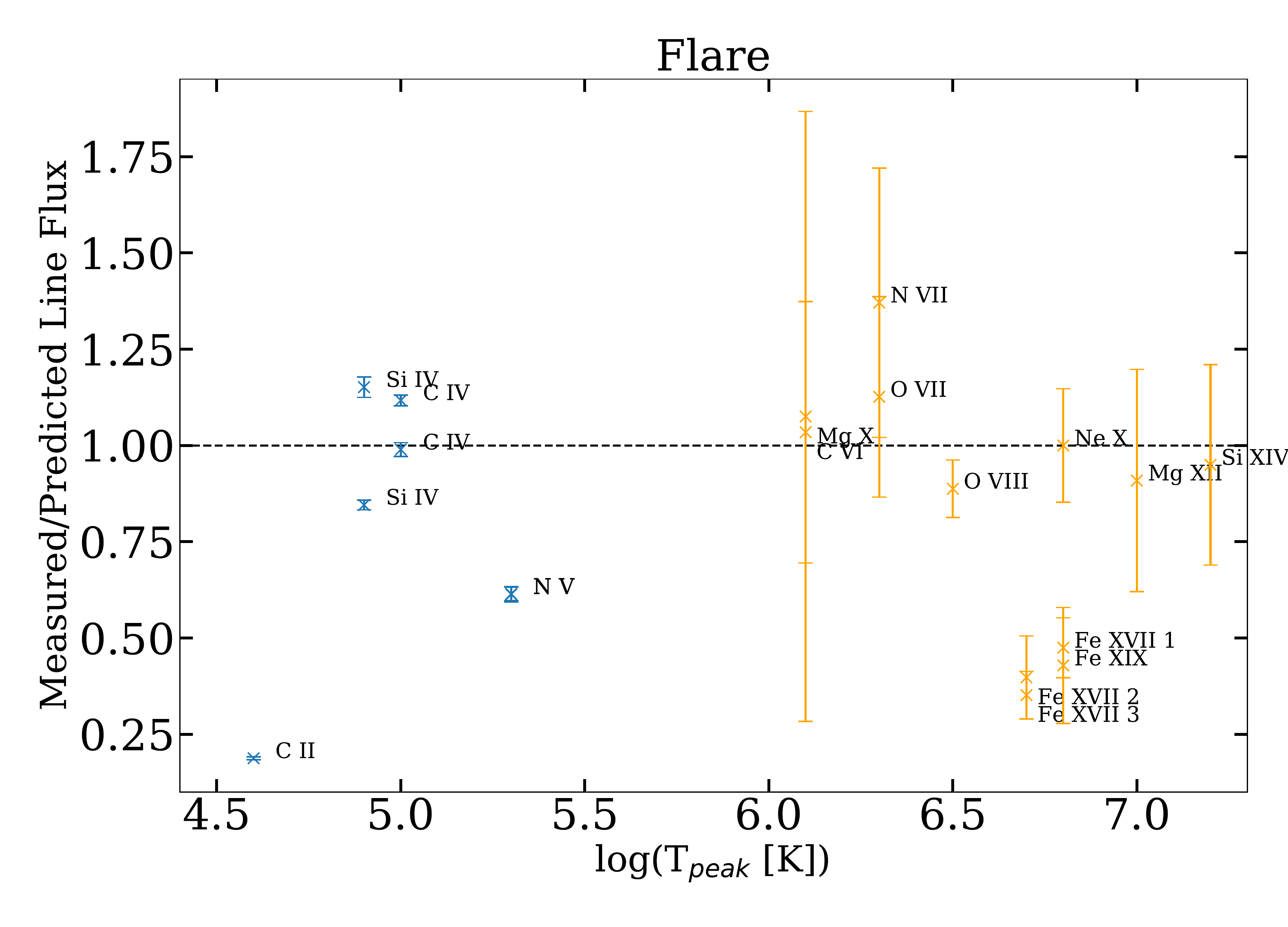}
    \caption{Flux ratios of measured to predicted quiescence line fluxes plotted against peak
        formation temperatures for quiescence (\emph{top}) and flaring (\emph{bottom}). The predicted fluxes were calculated
      with respective contribution functions and the determined DEM(T) distribution.
      }
    \label{fig:mp_qu}
\end{figure}

Moreover, we compare the line fluxes computed from this best-fit DEM function to the originally measured
line fluxes. 
In Fig. \ref{fig:mp_qu} 
 the ratios of measured to calculated fluxes of each emission line are plotted
during quiescence (top panel) and flaring (bottom panel) periods. Regarding quiescence, the DEM model predicts the line
fluxes well, except for the lines at highest peak formation temperature,
which also exhibit the largest line flux error. The typically
larger errors of the  X-ray predictions are caused by the lower data quality of the
\emph{Chandra} observations. For flaring periods, all emission lines are also predicted within 
an acceptable deviation. An extreme outlier is only \ion{C}{ii},
which is the line with the lowest peak formation temperature, which may indicate that
a larger order of the Chebyshev polynomial is still needed to reduce the
rigidity for the flaring DEM even further. Because optical depth effects may begin to play
a role here,
we  decided against increasing the order of the polynomial since
this moreover decreases the stability of the DEM fitting process. 
Furthermore, during the flaring state, all Fe lines are overpredicted, which
may hint at an overestimated abundancy during flare or a slightly overpredicted DEM at these temperatures, since all Fe lines are clustered in temperature.
Overall, the
DEM is expected to yield better predictions for quiescence since the longer quiescent
time spans allow one to accumulate more data than the shorter, though more intense, flaring periods.
Finally, to get an idea about how well the
DEM functions reflect the FUV and X-ray measurements and therefore the coronal (and transition
region) temperature structure of Proxima~Centauri, the flux ratio plot in Fig. \ref{fig:ratios}
is complemented with the ratio of the DEM functions (red line). The DEM structure
represents the flux ratio quite well, again with the exception of the very cool \ion{C}{ii}.
Since we do not have measurements of lines with a peak formation temperature between $\log$ T= 5.5-6.0,
we cannot exclude that there is a rise to higher flare to quiescent ratios in this regime. Nevertheless,
both X-ray and FUV data show a decrease of this ratio towards this temperature regime, making a trough-like
structure, as depicted in Fig. \ref{fig:ratios}, the most simple solution.

\subsection{Synthetic spectra}

\begin{figure}
    \centering
    \includegraphics[width=1\columnwidth]{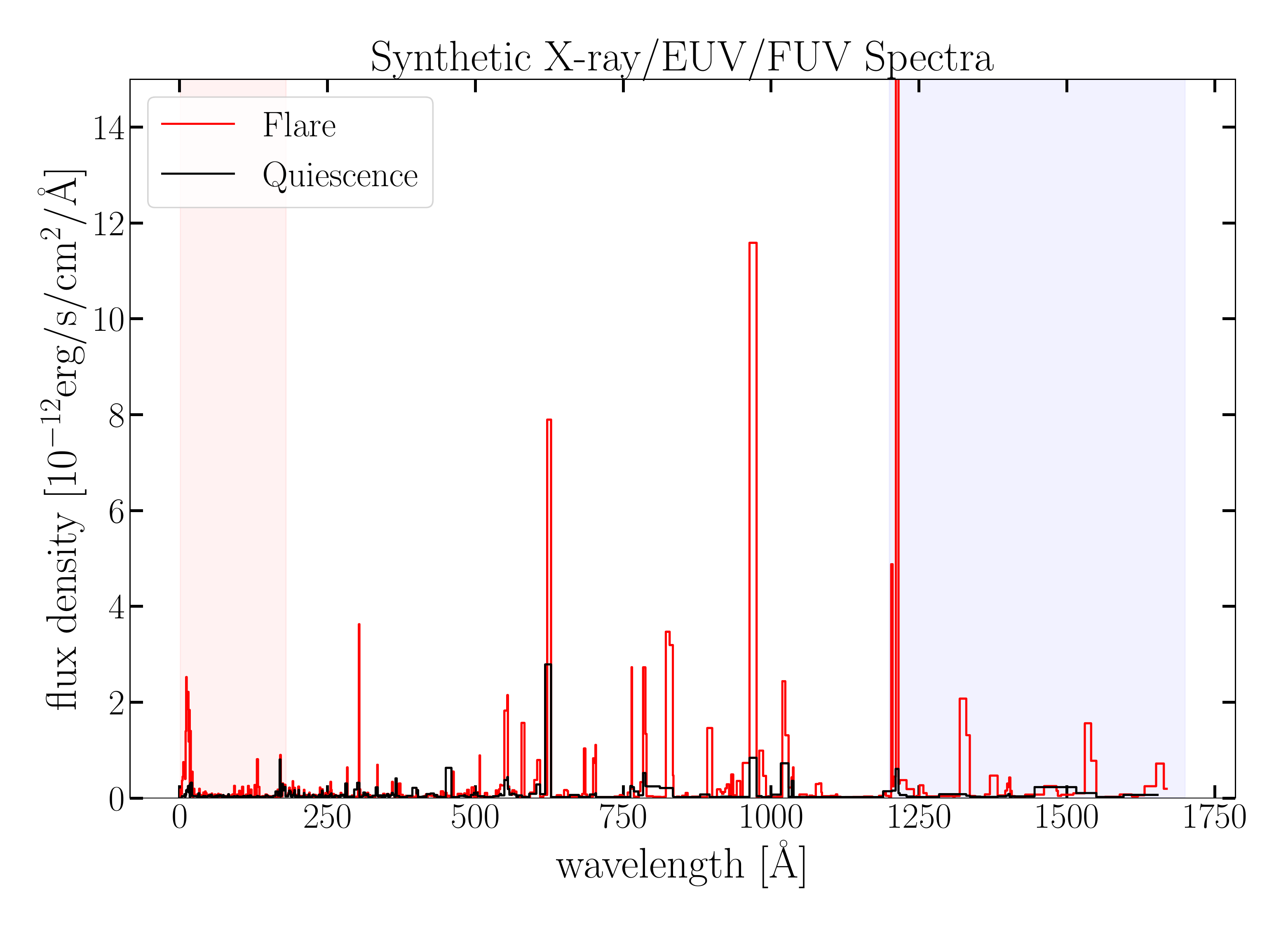}\\
    \includegraphics[width=1\columnwidth]{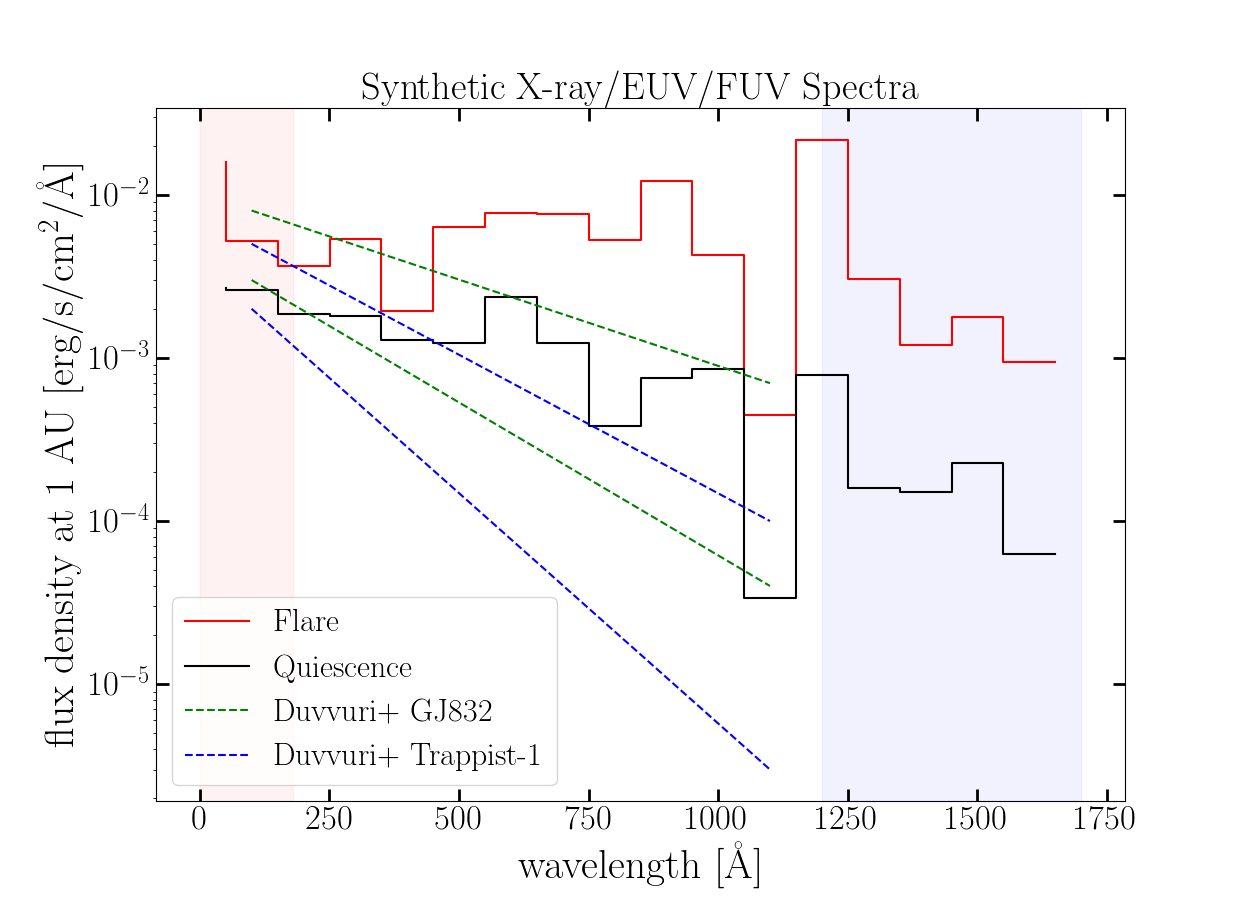}\\
    \caption{
    Synthetic spectra in the range of 1 to 1700 \AA, calculated from
        the DEMs for quiescent (black) and flaring (red)
        state of Proxima~Centauri. The pink shaded region marks the
        wavelength coverage of our \textit{Chandra} data, while the
        blue shaded region marks the coverage of the \textit{Hubble} data.
        For the flaring state, 
        the highest amplitude by far is
        seen for the Ly$\alpha$ line, which cannot be predicted well
        by our data since the line is formed at lower temperatures.
        The flux density was binned by a minimum flux of $2\cdot10^{-14}$ erg\,s$^{-1}$\,cm$^{-2}$\,\AA$^{-1}$ per bin
        (\emph{top}) and by 100\,\AA\, bins (\emph{bottom}) at a distance of 1\,AU.    }
    \label{fig:synspec}
\end{figure}

From the DEM of the quiescent and flaring state, we calculated synthetic spectra in the
range of 1 to 1700 \AA\ , respectively, using a resolution of 1 \AA. We show the synthetic spectra
in Fig. \ref{fig:synspec} with a wavelength binning defined by a minimum flux
density
of $2\cdot10^{-14}$ erg\,s$^{-1}$\,cm$^{-2}$\,\AA$^{-1}$ per bin.
Additionally, we show the flux density with a binning of 100\,\AA\, and at a distance
of 1\,AU to Proxima~Centauri for a better comparison to the synthetic spectra
constructed recently by \citet{Duvvuri2021} for four M dwarfs. 
Of these, we indicate the flux density covered by GJ~832 and Trappist-1 in 
Fig.\ref{fig:synspec}. 
Our quiescent synthetic spectrum of Proxima~Centauri shows 
a decline in the flux density in resemblance of their synthetic spectra of GJ~832 (M2/3) and 
Trappist-1 (M7.5), although the slope of our spectrum is somewhat flatter than for their spectra. 
Moreover, their synthetic
spectrum of Barnard's star (M4) during quiescence exhibits a slight increase in the flux density towards longer
wavelengths, while being relatively constant at a wavelength below 800 \AA. We find the slope
of Proxima~Centauri between those found for Barnard's star and GJ~832, but with flux
density values of Proxima~Centauri more closely resembling those of GJ~832. This better
agreement with an earlier spectral type may be explained by the higher activity level
of Proxima~Centauri compared to GJ~832, while Trappist-1 is again
much less active than Proxima~Centauri. 
Indeed, the flaring spectrum of Barnard's star also constructed by \citet{Duvvuri2021} 
is again in close resemblance to our flaring spectrum.

Further, we compared our synthetic spectrum to those constructed by
\citet{Loyd2016}. We cannot
reproduce the sharp drop at about 400\,\AA \, seen in all of their mid M dwarf spectra, and we note
that this feature is also absent in the spectra constructed by \citet{Duvvuri2021}. Yet
this feature is again seen in the composite spectrum of Proxima~Centauri reported by 
\citet{Ribas2017}, who used a variety of  observations obtained with different instruments. Otherwise 
our quiescent synthetic spectrum is also in agreement with these observations.

Our synthetic spectra can further be used for an evaluation of the stability of the
atmospheres of the two close-in planets in the system Proxima~Centauri~b and d 
\citep{Faria2022} as was demonstrated for Barnard's star by \citet{France2020}.
Since the mass loss is directly proportional to
the flux in the XUV range and our flare spectrum shows a factor of about 5 higher for the integrated
flux than the quiescent spectrum, even for our rather small flares,
the expected mass loss is also sensitively dependent on the flare
duty cycle of the star. 

For a crude estimate of the mass loss, 
we followed the example of \citet{Owen2012} and \citet{Poppenhaeger2021}, who
employed the widely used equation for energy limited hydrodynamic escape
$\dot{M}=\epsilon \frac{3\beta^{2}F_{XUV}}{4GK\rho_{pl}}$,
with $\epsilon$ being the efficiency of atmospheric loss, which we estimate
to be 0.1, with $K$ representing the impact of Roche overflow, which we estimate to
be 1 (i.e. no Roche overflow). Furthermore, we estimate the ratio between
the planetary radius and planetary FUV radius $\beta$ to be 1.5, since Proxima~Centauri
b and d are small planets. Finally, we estimate the planetary density $\rho_{pl}$
to be 5.0\,g/cm$^{3}$, which is slightly lower than the Earth's density. From our synthetic spectra,
we derived an integrated $F_{XUV}$ flux of 1.8\,erg/s/cm$^{2}$ during quiescent
intervals at the distance of Proxima~Centauri~b, and of 10.5\,erg/s/cm$^{2}$ during flaring
states, thus we estimate a present day mass loss of 0.4$\cdot 10^{9}$\,g\,s$^{-1}$ during
the quiescent and 2.1$\cdot 10^{9}$\,g\,s$^{-1}$ during flaring state, respectively.
Nevertheless, the detailed modelling of thermal and possibly non-thermal atmospheric escape induced by our
synthetic spectra on the two close-in planets 
is beyond the scope of this paper. 

\section{Conclusion}
\label{sec:conclusion}
In this work X-ray observations of Proxima Centauri taken by \emph{Chandra} and simultaneous
FUV spectra from \emph{Hubble/STIS}
were analysed. First, light curves were extracted, flaring as well as 
quiescent periods were determined, and associated spectra were generated. The spectra were used to measure
25 X-ray and 16 FUV emission lines out of which 12 and six, respectively, could be measured with
a high enough precision for a DEM construction. Flare to quiescence flux ratios range from about 1 for
example for \ion{Mg}{X} up to 
20 for \ion{Si}{IV}. The observed line fluxes were used to construct the DEM
of the optically thin coronal and transition region plasma.
The DEM functions yield the strongest 
increases at temperatures $\log$T=4.25-5.5 (20\,000-300\,000\,K, FUV regime) and beyond 
$\log$T=6.5 (3.1\,MK, X-ray regime) during flares. Furthermore, flare and quiescent line
fluxes were predicted with the help of the DEM curves and compared to measured values. For all 
but one very cool \ion{C}{ii} line, the model predicts acceptable values. The same methods as 
used in this work can be applied to many other spectra, in particular to archival Proxima 
Centauri data. The DEM method does not strictly require simultaneous X-ray and UV
observations, but it can be applied to X-ray and UV data individually, though each wavelength 
regime only would trace a certain range in plasma temperature. When using all suitable X-ray 
and UV observations, that is observations with sufficient spectral resolution and signal-to-noise,
one can determine how the spectral shape varies with time (or activity level) and construct 
the irradiation spectrum of Proxima~Centauri b for large ranges of conditions. These spectra can then 
be used to create sophisticated planetary atmosphere models, which will provide information about the 
conditions on Proxima~Centauri~b.

\bibliographystyle{aa} 
\bibliography{bib} 

\begin{appendix}
\section{Line lists}
Here we state the measured line properties of X-ray and FUV lines in Tables \ref{tab:xray_lines}
and \ref{tab:fuv_lines}.

\begin{table*}
\begin{center}
\caption{X-ray emission lines}
        \begin{tabular}{lccccc}
                \hline\hline
                Ion & Wavelength [$\mathring{A}$] & Net Line Flux [erg/s/cm$^2$] & Net Line Counts & Background Counts & log(T$_{\text{peak}}$ [K]) \\ 
                \hline
                \multicolumn{6}{c}{Quiescence} \\
                \hline
                Si XIV      &  6.186   &  9.25$\cdot 10^{-15}  \pm$ 1.45$\cdot 10^{-14}$ & 9.6 $\pm$ 15.1 & 109.33 $\pm$ 10.5 & 7.2\\
                Mg XII      &  8.419   &  1.14$\cdot 10^{-14}  \pm$ 1.15$\cdot 10^{-14}$ & 14.2 $\pm$  14.4 & 97.3 $\pm$ 9.9 & 7.0\\
                Ne X        &  12.138  &  1.03$\cdot 10^{-13}  \pm$ 1.38$\cdot 10^{-14}$ & 137.7  $\pm$ 18.5  & 102.3 $\pm$ 10.1 & 6.8   \\
                Fe XVII     &  15.013  &  1.21$\cdot 10^{-13}  \pm$ 1.24$\cdot 10^{-14}$ & 206.8  $\pm$ 21.2  & 121.6 $\pm$ 11.0 & 6.8   \\
                Fe XIX     &   15.208  &  3.37$\cdot 10^{-14}  \pm$ 1.00$\cdot 10^{-14}$ & 58.3 $\pm$ 17.4 & 121.7 $\pm$ 11.0 &  6.8\\
                Fe XVII     &  16.775  &  6.92$\cdot 10^{-14}  \pm$ 9.42$\cdot 10^{-15}$ & 132.0  $\pm$ 18.0  &  95.3 $\pm$  9.8 & 6.7   \\
                Fe XVII     &  17.096  &  1.23$\cdot 10^{-13}  \pm$ 1.21$\cdot 10^{-14}$ & 202.6  $\pm$ 19.8  &  95.3 $\pm$  9.8 & 6.7   \\
                O VIII      &  18.967  &  2.73$\cdot 10^{-13}  \pm$ 1.41$\cdot 10^{-14}$ & 519.7  $\pm$ 26.8  & 100.3 $\pm$ 10.0 & 6.5   \\
                O VII       &  21.602  &  1.26$\cdot 10^{-13}  \pm$ 1.42$\cdot 10^{-14}$ & 176.6  $\pm$ 19.9  & 109.6 $\pm$ 10.5 & 6.3   \\
                N VII       &  24.779  &  3.42$\cdot 10^{-14}  \pm$ 1.07$\cdot 10^{-14}$ & 52.8 $\pm$ 16.47 & 109.3 $\pm$ 10.5 & 6.3\\
                C VI        &  33.734  &  8.89$\cdot 10^{-14}  \pm$ 1.14$\cdot 10^{-14}$ & 140.7  $\pm$ 18.1  &  94.0 $\pm$  9.7 & 6.1   \\
                Mg X        &  57.876  &  2.04$\cdot 10^{-14}  \pm$ 4.22$\cdot 10^{-15}$ &  70.4  $\pm$ 14.6  &  70.6 $\pm$  8.4 & 6.1   \\
                \hline
                \multicolumn{6}{c}{Flare} \\
                \hline
                Si XIV      &  6.186   &  3.40$\cdot 10^{-13}  \pm$ 9.33$\cdot 10^{-14}$ & 27.1 $\pm$  7.4 & 14.0$\pm$ 3.7 & 7.2\\
                Mg XII      &  8.419   &  2.78$\cdot 10^{-13}  \pm$ 8.82$\cdot 10^{-14}$ & 26.4 $\pm$ 8.4 & 22.0 $\pm$ 4.7 & 7.0\\
                Ne X        &  12.138  &  8.05$\cdot 10^{-13}  \pm$ 1.18$\cdot 10^{-13}$ &   82.4 $\pm$ 12.1  &  32.3 $\pm$ 5.7 & 6.8 \\
                Fe XVII     &  15.013  &   5.06$\cdot 10^{-13} \pm$ 8.31$\cdot 10^{-14}$ &   65.9 $\pm$ 10.8  &  25.6$\pm$5.1 & 6.8   \\
                Fe XIX     &  15.208   &   1.87$\cdot 10^{-13} \pm$ 6.58$\cdot 10^{-14}$ &  24.8 $\pm$ 8.7 & 25.6 $\pm$ 5.1 & 6.8\\
                Fe XVII     &  16.775  &   2.02$\cdot 10^{-13} \pm$ 5.51$\cdot 10^{-14}$ &   29.7 $\pm$  8.1  &  17.6$\pm$4.2 & 6.7   \\
                Fe XVII     &  17.096  &   4.29$\cdot 10^{-13} \pm$ 7.52$\cdot 10^{-14}$ &   53.9 $\pm$  9.4  &  17.6$\pm$4.2 & 6.7   \\
                O VIII      &  18.967  &   1.15$\cdot 10^{-12} \pm$ 9.69$\cdot 10^{-14}$ &  167.1 $\pm$ 14.0  &  15.0$\pm$3.9 & 6.5   \\
                O VII       &  21.602  &   3.33$\cdot 10^{-13} \pm$ 7.69$\cdot 10^{-14}$ &   35.6 $\pm$  8.2  &  16.0$\pm$4.0  & 6.3   \\
                N VII       &  24.779  & 2.27 $\cdot 10^{-13} \pm$  5.79$\cdot 10^{-14}$ & 26.8 $\pm$ 6.8 & 10.0 $\pm$ 3.1 & 6.3 \\ 
                C VI        &  33.734  &   1.51$\cdot 10^{-13} \pm$ 4.98$\cdot 10^{-14}$ &   18.4 $\pm$  6.0  &   9.0$\pm$3.0  & 6.1   \\
                Mg X        &  57.876  &   2.26$\cdot 10^{-14} \pm$ 1.66$\cdot 10^{-14}$ &    6.0 $\pm$  4.4  &   6.6$\pm$2.6 & 6.1   \\
                \hline
        \end{tabular}
        \label{tab:xray_lines}
\end{center}
\end{table*}

\begin{table*}
\begin{center}
\caption{FUV emission lines}    
        \begin{tabular}{lccccc}
                \hline\hline
                Ion & Wavelength [$\mathring{A}$] & Net Line Flux [erg/s/cm$^2$] & Net Line Counts & Red-shift [km/s] & log(T$_{\text{peak}}$ [K]) \\
                \hline
                \multicolumn{6}{c}{Quiescence} \\
                \hline
                N V      & 1238.823 & 5.60$\cdot 10^{-14} \pm  1.03\cdot 10^{-15}$& 2782.7 $\pm$  51.1  & -22.8  & 5.3        \\
                N V      & 1242.806 & 2.81$\cdot 10^{-14} \pm  5.52\cdot 10^{-16}$& 2339.7 $\pm$  47.2  & -21.6  & 5.3        \\
                C II      & 1335.710 & 5.36$\cdot 10^{-14} \pm  6.42\cdot 10^{-16}$& 6292.7 $\pm$  72.7  & -23.2  & 4.6        \\
                Si IV     & 1393.757 & 2.35$\cdot 10^{-14} \pm  4.76\cdot 10^{-16}$& 2573.7 $\pm$  48.5  & -20.6  & 4.9        \\
                Si IV     & 1402.772 & 1.33$\cdot 10^{-14} \pm  4.56\cdot 10^{-16}$&  928.6 $\pm$  32.5  & -21.2  & 5.0        \\
                C IV      & 1548.189 & 1.54$\cdot 10^{-13} \pm  1.55\cdot 10^{-15}$& 8187.6 $\pm$  84.7  & -18.7  & 5.0        \\
                C IV      & 1550.775 & 8.25$\cdot 10^{-14} \pm  1.38\cdot 10^{-15}$& 3623.5 $\pm$  62.0  & -22.2  & 4.9        \\
                \hline
                \multicolumn{6}{c}{Flare} \\
                \hline
                N V      & 1238.823 & 2.28$\cdot 10^{-13} \pm  6.62\cdot 10^{-15}$& 1072.5 $\pm$  30.4  & -22.7  & 5.3        \\
                N V      & 1242.806 & 1.13$\cdot 10^{-13} \pm  3.63\cdot 10^{-15}$&  893.1 $\pm$  27.3  & -21.1  & 5.3        \\
                C II     & 1335.710 & 2.46$\cdot 10^{-13} \pm  4.75\cdot 10^{-15}$& 2731.7 $\pm$  55.7  & -21.6  & 4.6        \\
                Si IV    & 1393.757 & 3.86$\cdot 10^{-13} \pm  5.79\cdot 10^{-15}$& 3987.3 $\pm$  60.2  & -20.4  & 4.9        \\
                Si IV    & 1402.772 & 2.63$\cdot 10^{-13} \pm  6.00\cdot 10^{-15}$& 1726.1 $\pm$  41.5  & -20.4  & 5.0        \\
                C IV     & 1548.189 & 1.73$\cdot 10^{-12} \pm  2.16\cdot 10^{-14}$& 7444.6 $\pm$ 102.4  & -17.4  & 5.0        \\
                C IV     & 1550.775 & 7.69$\cdot 10^{-13} \pm  1.38\cdot 10^{-14}$& 3184.8 $\pm$  54.1  & -22.0  & 4.9        \\
                \hline
        \end{tabular}
        \label{tab:fuv_lines}
\end{center}    
\end{table*}

\end{appendix}

\end{document}